\documentclass[iop,revtex4,twocolappendix]{emulateapj}

\usepackage{txfonts}
\usepackage{textcomp}

\newcommand{\um}{\textmu m }
\newcommand{\uu}{\textmu m}
\newcommand{\ssil}{$S\!_{\rm{sil}}$ }
\newcommand{\ssilu}{$S\!_{\rm{sil}}$}
\newcommand{\neii}{[Ne {\sc ii}] }

\shorttitle{Mid-infrared spectral decomposition of AGN}
\shortauthors{Hern\'an-Caballero et al.}
\begin{document}
\title{Resolving the AGN and host emission in the mid-infrared using a model-independent spectral decomposition}
\author{
Antonio Hern\'an-Caballero,\altaffilmark{1}
Almudena Alonso-Herrero,\altaffilmark{1}
Evanthia Hatziminaoglou,\altaffilmark{2}
Henrik W. W. Spoon,\altaffilmark{3}
Cristina Ramos Almeida,\altaffilmark{4,5}
Tanio D\'iaz Santos,\altaffilmark{6,7}
Sebastian F. H\"onig,\altaffilmark{8}
Omaira Gonz\'alez-Mart\'in,\altaffilmark{9}
and Pilar Esquej\altaffilmark{10}
}
\email{ahernan@ifca.unican.es}
\altaffiltext{1}{Instituto de F\'isica de Cantabria, CSIC-UC, Avenida de los Castros s/n, 39005, Santander, Spain}
\altaffiltext{2}{European Southern Observatory, Karl-Schwarzschild-Str. 2, 85748 Garching bei M\"unchen, Germany}
\altaffiltext{3}{Cornell University, CRSR, Space Sciences Building, Ithaca, NY 14853, USA}
\altaffiltext{4}{Instituto de Astrof\'isica de Canarias, V\'ia L\'actea s/n, E-38205 La Laguna, Tenerife, Spain}
\altaffiltext{5}{Departamento de Astrof\'isica, Universidad de La Laguna, E-38206 La Laguna, Tenerife, Spain}
\altaffiltext{6}{N\'ucleo de Astronom\'ia de la Facultad de Ingenier\'ia, Universidad Diego Portales, Av. Ej\'ercito Libertador 441, Santiago, Chile}
\altaffiltext{7}{Spitzer Science Center, California Institute of Technology, MS 220-6, Pasadena, CA 91125, USA}
\altaffiltext{8}{School of Physics \& Astronomy, University of Southampton, Southampton SO18 1BJ, United Kingdom}
\altaffiltext{9}{Centro de Radioastronom\'ia y Astrof\'isica (CRyA-UNAM), 3-72 (Xangari), 8701, Morelia, Mexico}
\altaffiltext{10}{Departamento de Astrof\'isica, Facultad de CC. F\'isicas, Universidad Complutense de Madrid, E-28040 Madrid, Spain}

\begin{abstract}
We present results on the spectral decomposition of 118 \textit{Spitzer} Infrared Spectrograph (IRS) spectra from local active galactic nuclei (AGN) using a large set of \textit{Spitzer}/IRS spectra as templates. The templates are themselves IRS spectra from extreme cases where a single physical component (stellar, interstellar, or AGN) completely dominates the integrated mid-infrared emission. 
We show that a linear combination of one template for each physical component reproduces the observed IRS spectra of AGN hosts with unprecedented fidelity for a template fitting method, with no need to model extinction separately. We use full probability distribution functions to estimate expectation values and uncertainties for observables, and find that the decomposition results are robust against degeneracies.
Furthermore, we compare the AGN spectra derived from the spectral decomposition with sub-arcsecond resolution nuclear photometry and spectroscopy from ground-based observations. We find that the AGN component derived from the decomposition closely matches the nuclear spectrum, with a 1-$\sigma$ dispersion of 0.12 dex in luminosity and typical uncertainties of $\sim$0.19 in the spectral index and $\sim$0.1 in the silicate strength. We conclude that the emission from the host galaxy can be reliably removed from the IRS spectra of AGN. This allows for unbiased studies of the AGN emission in intermediate and high redshift galaxies --currently inaccesible to ground-based observations-- with archival \textit{Spitzer}/IRS data and in the future with the Mid-InfraRed Instrument of the James Webb Space Telescope. The decomposition code and templates are available at \url{http://www.denebola.org/ahc/deblendIRS}.
\end{abstract}


\section{Introduction} 

In the 5.5 years duration of the \textit{Spitzer} \citep{Werner04} cryogenic mission,   the Infrared Spectrograph \citep[IRS;][]{Houck04} obtained mid-infrared (MIR) spectra with unprecedented sensitivity for more than one thousand AGN \citep{Lebouteiller11}, including quasars, Seyferts, LINERs, radio-galaxies, blazars, etc.  
However, the limited spatial resolution of IRS observations (3.6" and 10.5" for the Short-Low (SL) and Long-Low (LL) modules, respectively) implies that the extended emission from the host galaxy usually contributes a significant fraction of the flux within the IRS aperture. Because of the difficulty in disentangling the emission from the AGN and its host galaxy, the shape of the AGN spectrum at MIR wavelenghts usually has large uncertainties, except for some quasar-luminosity AGN \citep{Mullaney11,Shang11,Shi13}.

\begin{figure} 
\includegraphics[height=6.0cm,width=8.4cm]{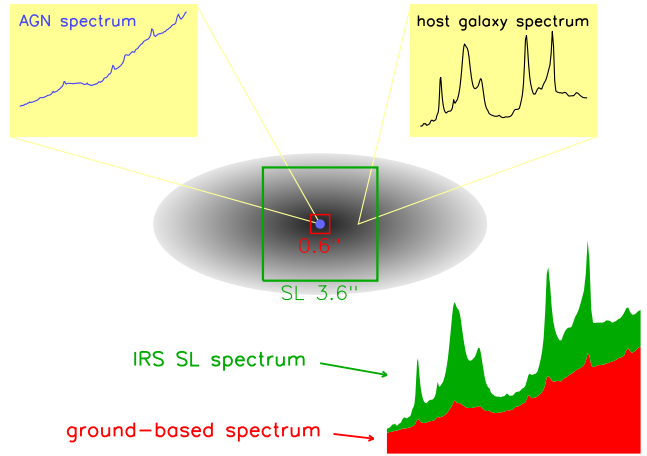}
\caption[]{Sketch illustrating the influence of spatial resolution in the spectra of local AGN. The red box represents the aperture size of a typical ground-based nuclear spectrum, while the green one represents the much larger aperture of the Spitzer/IRS Short-Low module. The spectrum observed by IRS is the sum of the ground-based spectrum and the spectrum in the area within the red and green squares, which only contains emission from the host.\label{fig:sketch}}
\end{figure}

Sub-arcsecond resolution observations with ground-based MIR spectrographs on 8-m and 10-m class telescopes probe nuclear regions in scales $<$100 pc for local AGN, effectively resolving away most of the emission from the host (see Figure \ref{fig:sketch}). This is  evidenced by the very small equivalent widths (EW) of the PAH features in the nuclear spectra \citep{Roche06,Honig10,Gonzalez-Martin13,Esquej14,Alonso-Herrero14,RamosAlmeida14} and the tight correlation between the MIR luminosity of the nuclear point source and the bolometric luminosity of the AGN derived from hard X-ray observations \citep[e.g.][]{Krabbe01,Horst08,Gandhi09,Asmus14}. 
However, ground-based MIR observations are limited to specific windows of atmospheric transmission and suffer from much lower sensitivity. Current samples of AGN with ground-based MIR observations are limited to a few dozens of local sources with $\sim$8--13\um spectroscopy \citep[e.g.][]{Gonzalez-Martin13,Esquej14} and a few hundreds with N-band and/or Q-band photometry \citep{Asmus14}. 

An alternative approach is to separate the spectral components corresponding to emission from the AGN and the host in the integrated spectrum. Multiple spectral decomposition methods have been applied to MIR spectra \citep[e.g.][]{Laurent00,Tran01,Lutz04,Sajina07,Hernan-Caballero09,Alonso-Herrero12a}, broadband spectral energy distributions (SEDs) \citep[e.g.][]{Hatziminaoglou08,Mateos15}, and combinations of both \citep[e.g.][]{Mullaney11,Feltre13}.

Libraries of theoretical models abound in the literature both for the IR emission of AGN
\citep[e.g.][]{Pier92,Pier93,Granato94,Efstathiou95,Fritz06,Nenkova08} and galaxies \citep[e.g.][]{Silva98,Popescu00,Gordon01,Siebenmorgen07,Draine07,daCunha08,Galliano08}. 
While they are suitable for fitting spectral energy distributions spanning a wide wavelength range, in general they cannot reproduce accurately, for example, the diverse shapes of the 10\um silicate feature  
or the relative strength of individual PAH bands.
More detailed models of the MIR spectrum containing dozens of physically motivated components, such as those produced by CAFE \citep{Marshall07} or PAHFIT \citep{Smith07} and its variants \citep[e.g.][]{Gallimore10}, allow to reproduce these features with high fidelity. However, the large number of free parameters involved (more than 100 in the case of PAHFIT) implies that solutions are often degenerate except in very high signal-to-noise ratio (SNR) spectra.

Another approach that requires few free parameters yet provides acceptable fits for many MIR spectra is spectral decomposition with empirical templates.
Most previous attempts at decomposing MIR spectra with empirical templates have relied on two or three spectral components represented by universal templates \citep[e.g.][]{Laurent00,Tran01,Nardini08,Hernan-Caballero09,Hernan-Caballero11,Veilleux09,Ruiz13}. This technique usually models the effects of extinction by assuming a dust screen with a particular extinction law that applies to at least the AGN component, and uses the value of $A_V$ as an extra free parameter.
This simple approach has a caveat: while the MIR emission from the host galaxy is relatively uniform across galaxies, it is simply not possible to reproduce the full range of SEDs observed in AGN-dominated spectra by just varying the amount of extinction applied to a single AGN template. This is not critical when only a narrow spectral range is considered \citep[e.g.][]{Nardini08,Ruiz13}, but it quickly becomes an issue as the fitted spectral range grows. For example, \citet{Hernan-Caballero09} showed that AGN with a steeper 5--15\um slope compared to the chosen AGN template require increasingly high contributions from a starburst continuum (HII region) template to obtain acceptable fits, even if no signs of star formation (as in PAH emission) are observed.

A complete description of the spectral components in a given spectrum requires not only finding the parameters of the best fitting model, but also their uncertainties. Uncertainties are caused by both noise in the data and degeneracies in the template set. While most template fitting routines account for the former, the latter are usually ignored. 
In the last few years, algorithms based on Bayesian inference have been developed to find the marginalised probability distribution of physical parameters from comparison of the observed data with all the models in the library. Some remarkable examples are MAGPHYS \citep{daCunha08} and BayesCLUMPY \citep{AsensioRamos09}. However, all these methods rely on grids of theoretical models with a uniform sampling of the parameter space to facilitate the calculation of probability distributions. Recently, \citet{Noll09} proposed a new Bayesian-like marginalisation method that is suitable for models with an irregular coverage of the parameter space, laying the groundwork for a complete description of spectral components based on purely empirical templates.
 
In this paper we take a new approach to spectral decomposition that uses libraries of observed high SNR spectra as templates. We show that the method produces excellent fits that accurately reproduce most features in high SNR spectra, yet it is robust against degeneracies even in poor SNR spectra. We obtain reliable expectation values and uncertainties for AGN properties by computing full probability distribution functions (PDFs) using a marginalisation method that overcomes the inhomogeneity in the template set. 
By comparing the derived AGN properties with those obtained from ground-based high spatial resolution (0.4"--0.8") MIR photometry and spectroscopy, which exclude circumnuclear emission, we demonstrate that this kind of spectral decomposition is accurate at separating the emission from the AGN and its host galaxy and recovering the shape of the AGN spectrum.
In \S2 we present the method of spectral decomposition and our approach to the calculation of the expectation values of observables. \S3 describes the set of templates and \S4 the selection of the test sample. In \S5 we present the results of the spectral decomposition and perform a careful validation by comparing the AGN fluxes, spectral indices, and silicate strengths with those obtained from ground-based MIR photometry and spectroscopy. \S6 discuses the strengths and limitations of the method in light of the presented results, and \S7 summarizes our main conclusions. Throughout this paper we use a cosmology with $H_0$ = 70 km s$^{-1}$ Mpc$^{-1}$, $\Omega_M$ = 0.3, and $\Omega_\Lambda$ = 0.7.

\section{The method}\label{themethod}

Our approach to spectral decomposition relies on the assumption that the spectral shape of the AGN and host galaxy emissions that contribute to the integrated spectrum can also be found in other sources where either the AGN or the host completely dominates the emission.
The emission from the host is itself a combination of that from the stars and the interstellar medium, with the former dominating in passively evolving galaxies (at least at $\lambda$$<$15\uu) and the latter in star-forming ones. 
Accordingly, any IRS spectrum could be reproduced by a linear combination of three IRS spectra selected from sets containing only extreme cases of `pure-AGN', `pure-stellar', and `pure-interstellar' spectra. For clarity, we refer to these single-component spectra as templates.

This procedure could raise several concerns. One is that AGN feedback could have an impact on the interstellar medium, modifying the spectrum of the host, e.g., by suppressing the PAH features. However, nuclear spectra of nearby AGN hosts at high spatial resolutions show no clear sign of PAH suppression, even in the close vicinity of the AGN \citep{Esquej14,Alonso-Herrero14,RamosAlmeida14,Garcia-Bernete15}. 
Another issue could be that the host galaxy modifies the intrinsic AGN spectrum via foreground extinction from dust in the line of sight to the AGN. While foreground extinction can indeed be very strong in luminous infrared galaxies (LIRGs), mergers, and edge-on galaxies \citep{Goulding09,Alonso-Herrero11,Gonzalez-Martin13}, we do not intend to recover the intrinsic spectrum of the AGN, but the one that would be observed at higher spatial resolution. 
Since the extinction caused by the host is independent on its angular size or the luminosity of the AGN, it should be possible to find a `pure-AGN' template with the same intrinsic spectrum and the same level of (host induced) extinction but negligible host emission.

Modelling foreground absorption in the MIR with an extinction law and a few empirical or semi-empirical `intrinsic' AGN templates provides great flexibility. However, this relies on the assumption of the extinction law and is a path already well explored \citep[e.g.][]{Tran01,Lutz04,Spoon04,Hernan-Caballero09,Mullaney11}. 
Our aim in this work is to produce spectral decompositions that are completely model-independent. Therefore, we require the template set to be large and varied enough to reproduce both the diversity in the intrinsic spectra of AGN and hosts \textit{and} the different levels of foreground extinction to be found. 

Our decomposition method is implemented in the DeblendIRS routine\footnote{Available at \url{http://www.denebola.org/ahc/deblendIRS}}, written in the IDL language.
We build a library of templates (see \S\ref{templates-sect}) containing IRS spectra of galaxies whose emission is strongly dominated by the stellar population (stellar templates), the interstellar medium (PAH templates), or the active galactic nucleus (AGN templates). 
All the spectra (the ones to be decomposed and the templates) are de-redshifted and resampled to a common wavelength grid $\lambda_n$. We choose to resample with a wavelength resolution $\Delta\lambda$=0.1\uu. While the results of the decomposition do not depend significantly on $\Delta\lambda$, this value is larger than the resolution of the original spectra and therefore increases the SNR per resolution element. The flux uncertainties $\sigma$($\lambda$) are calculated for the resampled spectra assuming that random errors for adjacent pixels are not correlated. 

For every galaxy in the sample we try spectral decompositions using every possible combination of one stellar template, one PAH template, and one AGN template. We represent each of these combinations with the tern of indices ($i$,$j$,$k$) for the $i$-th stellar template, the $j$-th PAH template, and the $k$-th AGN template.
The decomposition model for the spectrum F($\lambda$) with the tern ($i$,$j$,$k$) is then:

\begin{equation}
f_{i,j,k}(\lambda) = a f^{star}_i(\lambda) + b f^{PAH}_j(\lambda) + c f^{AGN}_k(\lambda)
\end{equation}

\noindent where $f^{star}_i(\lambda)$, $f^{PAH}_j(\lambda)$, and $f^{AGN}_k(\lambda)$ are the restframe flux densities of the templates, and the coefficients $a$, $b$, $c$ are obtained by minimising the $\chi^2$:

\begin{equation}
\chi^2 = \sum_{n=1}^{N} \frac{(F(\lambda_n) - f_{i,j,k}(\lambda_n))^2}{\sigma^2(\lambda_n)}
\end{equation}
 
Since the model is linear on the coefficients, we can find the values that minimise $\chi^2$ algebraically. This is much faster than an iterative fit and allows to exhaustively explore all the possible template combinations in a few seconds. If any of the fit coefficients turns out to be negative, we fix it at zero and repeat the calculation for the other two until we arrive at a solution where all the coefficients are non-negative.

We compare the minimum $\chi^2$ for each combination of templates to determine the one that produces the best $\chi^2$ of all models. 
We note that while the absolute minimum of $\chi^2$ identifies the best fitting model, its value is not a reliable indicator of the similarity between the model and the actual spectrum of the source, since it also depends on the SNR of the spectrum \citep[see e.g.][]{Hernan-Caballero12}. At high SNR, small differences between the spectrum and the model can be statistically significant given the small photometric errors, and it is not unusual to obtain a $\chi^2$ per degree of freedom $\chi^2_\nu$ = $\chi^2$/(N-3) $\gg$ 1. However, in lower SNR spectra differences between the spectrum and the model are blurred by the noise, and $\chi^2_\nu$$\lesssim$1 is typical. 
 
A complementary statistic on the differences between the model and the spectrum is the coefficient of variation of the root-mean-square error (CV$_{\rm{RMSE}}$):

\begin{equation}
CV_{\rm{RMSE}} = \frac{\sqrt{\sum_{n=1}^N \big{(}F(\lambda_n) - f_{i,j,k}(\lambda_n)\big{)}^2/N}}{\sum_{n=1}^N F(\lambda_n)/N}
\end{equation}

\noindent which represents the ratio of the typical residual to the mean flux. This is formally equivalent to a $\chi^2_\nu$ that assumes flux uncertainties are proportional to fluxes, and therefore ignores the SNR of the spectra. Accordingly, a comparison of the two statistics provides a reliable diagnostics of the quality of the fit irrespective of the SNR of the spectrum.

At constant SNR, both $\chi^2_\nu$ and CV$_{\rm{RMSE}}$ decrease for increasingly accurate fits. However, unlike $\chi^2_\nu$, CV$_{\rm{RMSE}}$ increases if the SNR is degraded. This is because larger flux uncertainties increase the residuals (the numerator in Eq. 3), but not the average flux.
Obtaining high values for both CV$_{\rm{RMSE}}$ and $\chi^2_\nu$ implies that no model can correctly reproduce the spectra, while low $\chi^2_\nu$ and CV$_{\rm{RMSE}}$ indicates a good fit with small relative residuals in a high SNR spectrum. However, a low $\chi^2_\nu$ with high CV$_{\rm{RMSE}}$ implies the model is compatible with the data just because the spectrum is noisy.
While $\chi^2_\nu$ values are not directly comparable among sources due to differences in the SNR of their spectra, the high SNR of all the templates ensures that for a given source we can arrange the models by the quality of their fit using $\chi^2_\nu$ alone.

The empirical nature of the templates implies that we cannot expect a homogeneous and complete sampling of the parameter space defined by the physical properties of interest (such as the spectral index of the AGN component, $\alpha$, and the strength of its silicate feature, \ssilu). Instead, the most common values are well represented in the library by many templates that are similar to each other, while extreme values are underrepresented.
Since the probability distribution function (PDF) of a physical property is usually computed as the sum of the probabilities of all the models in given bins of the parameter space, the inhomogeneous sampling acts as a prior that favours the most densely sampled regions of the parameter space.
Thefore, to obtain reliable PDFs we need a method that is robust against variations in the density of templates throughout the parameter space, which would otherwise favour values that are overrepresented in the library. One such method is the `max' method described in \citet{Noll09}.
Very briefly, we take the entire range of possible values for the parameter of interest, $x$, and split it into same-sized bins. For each bin, $i$, we find all the models that produce a value for the parameter within its limits, and take the lowest $\chi^2$, $\chi^2_i$. 
The probability of the parameter $x$ to take a value inside the bin $i$ is then:

\begin{equation}
p_i = \frac{e^{-\chi^2_i/2}}{\sum e^{-\chi^2_i/2}}
\end{equation}

\noindent where the normalisation ensures that the sum of the probabilities for all the bins is 1.
Then, the expectation value for $x$ is:

\begin{equation}
\langle x \rangle = \sum p_i x_i
\end{equation}

\noindent and its standard deviation is:

\begin{equation}
\sigma_x = \sqrt{\sum p_i (x_i - \langle x \rangle)^2} 
\end{equation}

Obtaining a realistic PDF with the `max' method requires the bins to be large enough for each of them to contain at least one model. This is guaranteed for properties that depend on the \textit{coefficients} of the three spectral components, such as the fractional contribution of the AGN to the restframe 12\um luminosity (L$^{\rm{AGN}}_{\rm{12}}$), given the large number of possible template combinations. However, for properties that depend on the \textit{templates} themselves, such as $\alpha$ and \ssilu, obtaining a correct sampling depends on the library.

We restrict the modelling of the IRS spectrum to the restframe 5--15\um spectral range.
This is because the most prominent PAH features (which are essential to disentangle AGN and host emission) reside in this range, and at $\lambda$$>$15\um discriminating between the continuum emission of the AGN and the host becomes increasingly difficult. In addition, the fraction of the total continuum emission arising from the AGN decreases at longer wavelengths, due to the combined effects of a flat AGN spectrum and a steeply rising continuum from dust heated by star formation \citep[see e.g.][]{Mullaney11}. In extended sources, this effect is exacerbated by the the larger aperture of the Long-Low (LL; 14--38\uu; 10.5") IRS module compared to Short-Low (SL; 5.2--14\uu; 3.6") which increases the contamination from the host in the latter.

Setting the long wavelength limit at 15\um restframe also implies that we can use as AGN templates sources at redshifts up to $z$=1.5 while mantaining full spectral coverage. This is important because the diversity in AGN spectral shapes requires the library of `pure-AGN' spectra to be as large as possible for the best results (see \S\ref{templates-sect}).

\section{Spectral templates}\label{templates-sect}

\begin{figure*} 
\includegraphics[height=6.0cm,width=8.4cm]{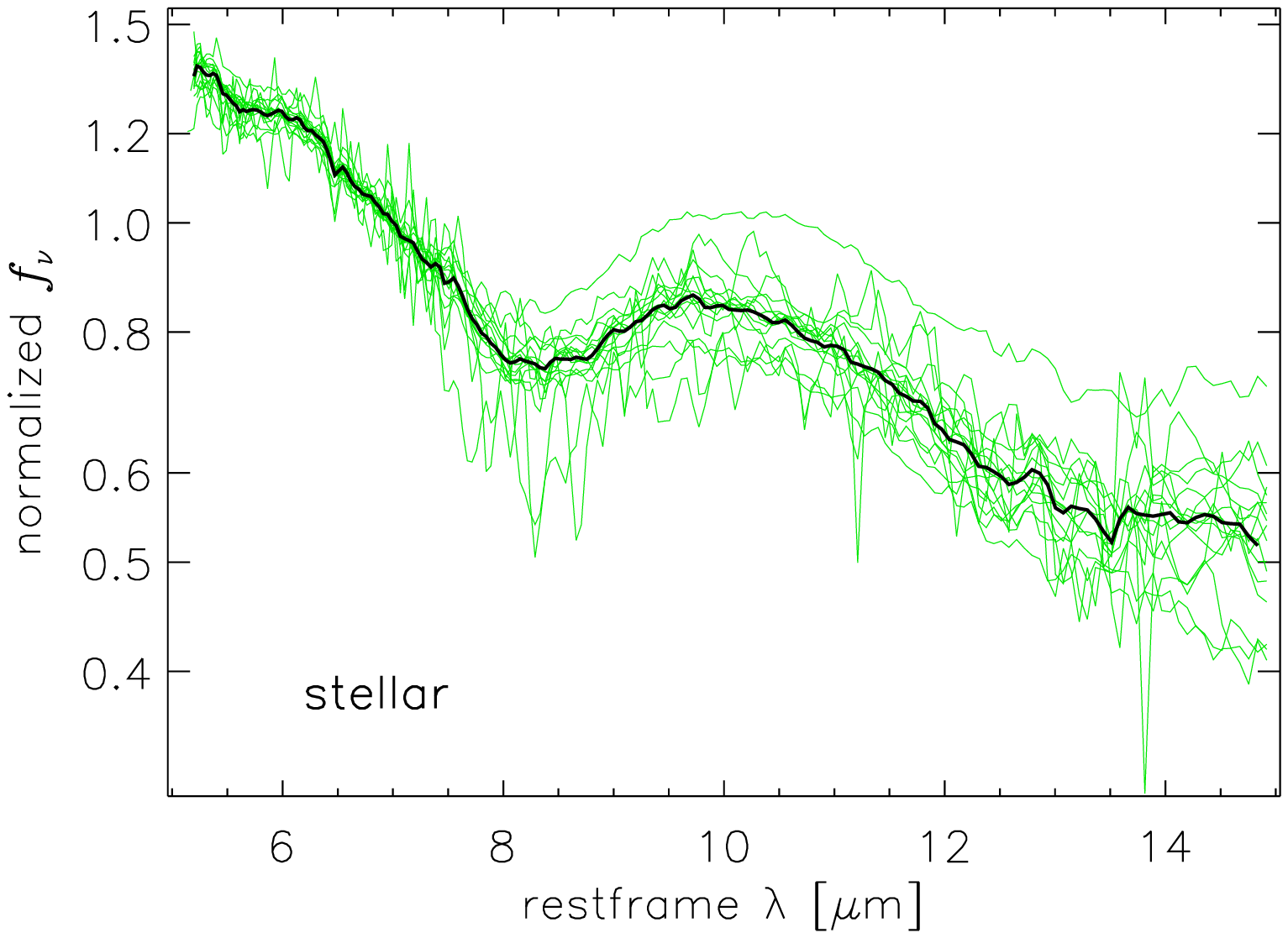}\hfill
\includegraphics[height=6.0cm,width=8.4cm]{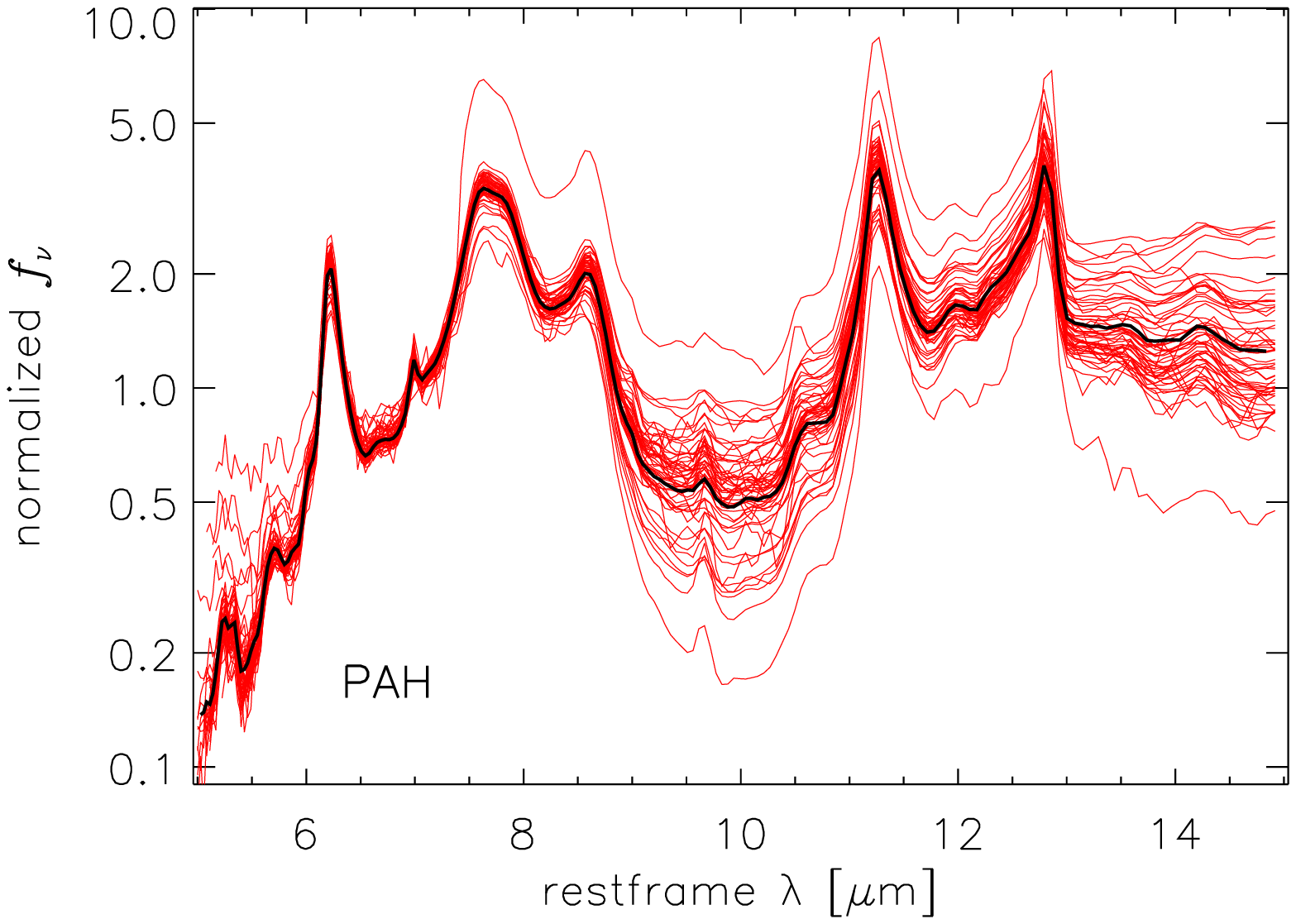}
\includegraphics[height=6.0cm,width=8.4cm]{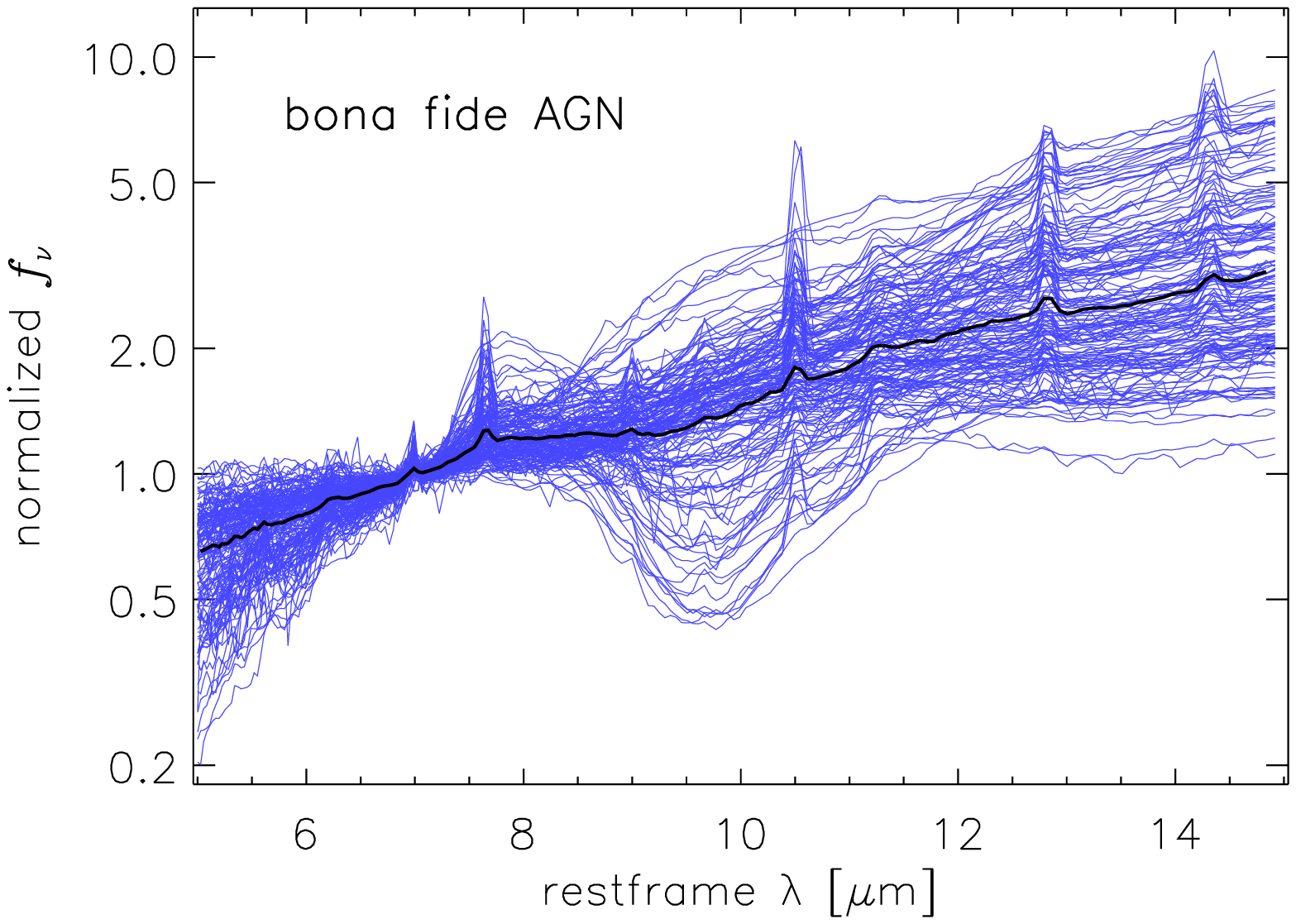}\hfill
\includegraphics[height=6.0cm,width=8.4cm]{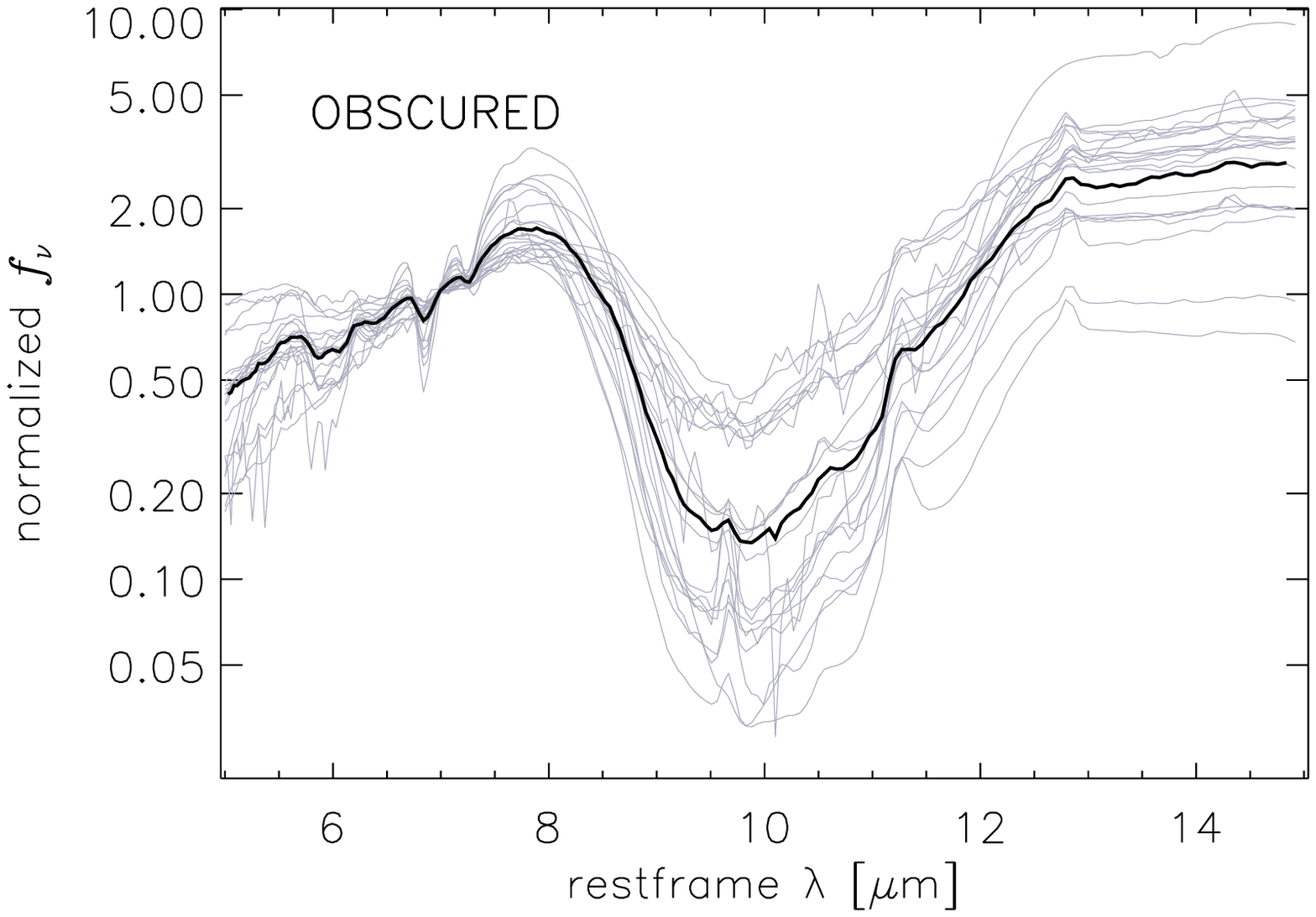}
\caption[]{The entire set of templates normalised at 7\um restframe and grouped by categories (see text for details). The thick solid line represents the average spectrum for each group.\label{fig:templates}}
\end{figure*}

All the templates employed here are IRS low-resolution spectra from CASSIS of individual galaxies with full spectral coverage in the restframe range 5.3--15.8\uu, reliable spectroscopic redshift, and high SNR. All the templates have median SNR per pixel $>$10 in the original spectra and $>$15 after the resampling. Typical values are much higher, with 75\% of templates at SNR/pixel$>$100 after resampling. 

We select templates of three types, each of them representing a distinct physical component of the integrated MIR emission of galaxies: the stellar population (stellar templates), the interstellar medium (PAH templates) and the AGN (AGN templates). We put great care into making sure that only one of these components contributes significantly to the observed spectrum of the galaxies selected as templates. For this, we rely on three criteria: the slope of the MIR continuum, the strength of the PAH features, and the optical spectroscopic classification.

For the `stellar' templates, we select 19 local elliptical and S0 galaxies. To ensure they have negligible star formation, we require the PAH bands to be very weak or absent, with equivalent widths for the 6.2\um (EW$_{62}$) and 11.3\um (EW$_{113}$) PAH bands $<$0.02\um \citep[see][for the details on the measurement of the EW of the PAH bands]{Hernan-Caballero11}. We also check that the IRS spectra have a blue stellar-like MIR continuum and the sources are not classified as AGN in the NASA Extragalactic Database\footnote{http://ned.ipac.caltech.edu} (NED). 

The 56 `PAH' templates are IRS spectra of normal star-foming and starburst galaxies at redshifts up to $z$=0.14. We make sure these sources do not have significant stellar contributions to their MIR spectra by requiring both high EW of the PAH features (EW$_{62}$ $>$ 1.0\um and EW$_{113}$ $>$ 1.0\uu) and a very weak continuum at 5\uu. We also verify that they are not classified as AGN in NED.

Finally, the AGN templates are IRS spectra of sources classified in the optical as quasars, Seyfert galaxies, LINERs, and blazars. We also include a variety of optically obscured AGN and radiogalaxies. The library of AGN templates includes sources at redshifts from $z$=0.002 to $z$=1.4 and covers several orders of magnitude in bolometric luminosity. 
We ensure that the AGN templates do not contain any significant emission from the host galaxy by requiring the PAH features to be extremely weak or absent (EW$_{62}$ $<$ 0.02\um and EW$_{113}$ $<$ 0.02\uu). 

Some of the AGN templates are from (U-)LIRGs with a deeply obscured nucleus, as evidenced by the strong water ice absorption bands and a deep silicate feature, together with very weak or absent PAH features. In these sources, the observed spectrum depends mostly on the geometry of the dust distribution, while the nature of the power source (either AGN, starburst, or a combination of both) plays only a minor role \citep{Levenson07,Imanishi07}. As a consequence, we cannot evaluate the actual AGN contribution to the MIR emission in these sources.

We inspect all the AGN templates with the 10\um silicate feature in absorption and classify them as bona fide AGN (AGN) or obscured nuclei (OBSCURED) depending on the absence/presence of H$_{\rm{2}}$O absorption bands at $\sim$6\um and/or a silicate strength \ssil$<$-1.5 (see \S\ref{AGNproperties_subsect} for the definition of \ssilu).
Removing the OBSCURED templates from the set of AGN templates would force the decomposition routine to choose AGN templates that provide poor fits to obscured sources. Therefore, we opt to keep the OBSCURED templates, but consider as upper limits the AGN fractions and AGN luminosities obtained in cases where the best fitting model involves an OBSCURED template.
The final selection of AGN templates contains 181 spectra (159 bona fide AGN and 22 OBSCURED sources). 

Figure \ref{fig:templates} shows all the templates selected for use in the spectral decomposition, separated by type. Bona fide AGN and OBSCURED sources are shown separately for clarity. Spectral shapes show very small source to source variations among stellar templates, as is expected for spectra dominated by the Rayleigh-Jeans tail of stellar emission. PAH and OBSCURED templates show also small variations, mostly caused by the strength of the PAH and silicate features, respectively. The largest diversity of shapes is found among the bona fide AGN templates. These show a wide range of spectral slopes, silicate feature strengths, and intensity of fine structure lines.

\section{Sample selection}

To demonstrate the power of our decomposition method we select a sample of local active galaxies for which there is available \textit{Spitzer}/IRS spectroscopy as well as ground-based MIR high angular resolution imaging. 

Our parent sample is drawn from the subarcsecond-resolution MIR imaging atlas of local AGN presented in \citet{Asmus14}. The atlas includes all the publicly available (by early 2014) MIR imaging with ground-based 8-m class telescopes. A total of 895 independent photometric measurements with a variety of filters in the N and Q atmospheric windows are presented for 253 distinct sources, with a median redshift of $z$=0.016.
 
We performed a search by coordinates in the Cornell Atlas of \textit{Spitzer}/IRS Sources\footnote{http://cassis.astro.cornell.edu} \citep[CASSIS;][]{Lebouteiller11} version 6.  We used a 10" search radius and found IRS spectra for 157 of the 253 sources.
We verified that coordinates of the nuclear point-source derived from ground-based observations are within the IRS Short-Low aperture in all cases. These 157 sources comprise our main working sample. Their redshift distribution ranges from $z$=0.00024 to $z$=0.23, with a median value of $z$=0.017. 

\citet{Asmus14} provide optical spectroscopic classifications for the sources in their sample. Our subsample contains 46 type 1 AGN (Sy1, Sy1.2, Sy1.5), 8 intermediate type AGN (Sy1.8, Sy1.9), 64 type 2 AGN (Sy2), 13 low-ionization nuclear emission line regions (LINER), 24 starburst/AGN composites (Cp) and 2 additional sources with uncertain classifications.

Ground based N-band spectroscopy ($\sim$8--13\uu) with spectral resolutions between 0.4" and 0.8" is also available for 28 of our sources from the sample of local Seyferts compiled by \citet{Esquej14}. The observations where originally published by 
\citet[][ Circinus]{Roche06},
\citet[][ NGC 5506, NGC 7172]{Roche07},
\citet[][ IC 5063]{Young07},
\citet[][ NGC 5135, NGC 7130]{Diaz-Santos10},
\citet[][ ESO 323-G77, Mrk 509, NGC 2110, NGC 3227, NGC 3783, NGC 4507, NGC 7213, NGC 7469]{Honig10},
\citet[][ NGC 4151]{Alonso-Herrero11},
\citet[][ NGC 2992, NGC 4388, NGC 5347]{Colling11},
\citet[][ NGC 1365]{Alonso-Herrero12b},
\citet[][ NGC 1386, NGC 3081, NGC 4945, NGC 5128, NGC 5643, NGC 7479, NGC 7582]{Gonzalez-Martin13}, and
\citet[][ NGC 1808, NGC 3281]{Sales13}.

\section{Results}

\subsection{Spectral decomposition models}

We obtain spectral decompositions for the 118 sources with ground based sub-arcsec resolution MIR photometry and \textit{Spitzer}/IRS spectra that were not selected as AGN templates. For the remaining 39 sources, their best fitting model is trivially one with themselves as the AGN template and 0\% PAH and stellar components. We confirm that the PAH and stellar components are actually negligible in these sources by temporarily removing their spectra from the template library and finding their new best fitting model. The `max' method allows to obtain expectaction values and uncertainties for physical parameters also for these sources.

The parameters of the best decomposition model for all the sources in the sample are listed in Table \ref{table:decomposition}. 
Columns 3 to 8 identify the PAH, AGN, and stellar templates involved in the best fitting model, and list the fractional contributions of each of them to the
integrated 5--15 \um luminosity.
Columns 9 and 10 indicate the reduced $\chi^2$ of the model, $\chi^2_\nu$, and the coefficient of variation of the root-mean-square error, CV$_{\rm{RMSE}}$. 
For the 118 sources that are not templates themselves, values of $\chi^2_\nu$ are between 0.16 and 126 (median value: 2.17). Values of CV$_{\rm{RMSE}}$ range from 0.013 to 0.51 with a median of 0.048. The latter implies that typical fitting residuals are $\sim$5\% of the flux density of the spectra, equivalent to the residuals that would be obtained with a `perfect' model in spectra with SNR/pixel=20 after resampling.

\begin{figure} 
\includegraphics[width=8.4cm]{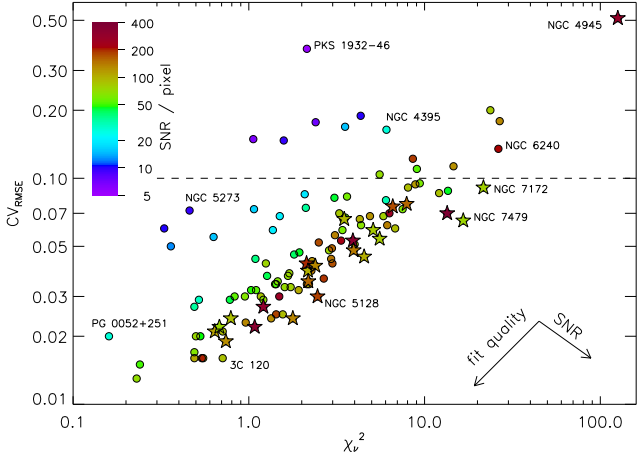}
\caption[]{Comparison between the coefficient of variation of the root-mean-square error and the reduced $\chi^2$ in the best-fitting model for each source. Stars represent sources with ground-based MIR photometry and spectroscopy, while circles represent those with just photometry. The colour coding indicates the median signal to noise ratio of the resampled IRS spectra. The dashed line marks the CV$_{\rm{RMSE}}$=0.1 threshold used to distinguish good and poor fits.\label{fig:CV-chi2}}
\end{figure}

Figure \ref{fig:CV-chi2} represents CV$_{\rm{RMSE}}$ versus $\chi^2_{\nu}$ for the best fitting decomposition model of each source. The sources with the highest SNR in their IRS spectra ($\gtrsim$50 after resampling) concentrate in a diagonal sequence, while all the sources with SNR$<$50 are found above the sequence, at increasing distances with decreasing SNR. Because of the sensitivity constraints for ground-based spectroscopy, sources in this subsample have on average higher fluxes and SNR in their IRS spectra compared to the general sample, but their CV$_{\rm{RMSE}}$ and $\chi^2_{\nu}$ are similar to those from other sources with the same SNR.

Visual inspection of the fits confirms that the accuracy of the decomposition model decreases along the diagonal sequence, from the bottom left to the top right corners of the diagram. The best models, with $\chi^2_{\nu}$$<$1 and residuals of $\lesssim$2\%, are barely discernible from the spectra they replicate. On the other hand, the sources with the highest values of $\chi^2_{\nu}$ ($\gtrsim$10) have large residuals in their fits, indicating that their high $\chi^2_{\nu}$ is not due to an underestimation of the photometric errors but caused by the lack of suitable templates. The most extreme case is NGC 4945, whose peculiar spectrum is dominated by a deeply obscured starburst component that is outside the range of variation of our PAH templates.
There are a few sources far from the sequence with $\chi^2_{\nu}$$<$10 and CV$_{\rm{RMSE}}$$>$0.1. All these spectra have much lower SNR compared to those in the sequence. Their larger flux uncertainties explain their relatively moderate $\chi^2_{\nu}$ values in spite of the large residuals.

\begin{figure} 
\includegraphics[width=8.4cm]{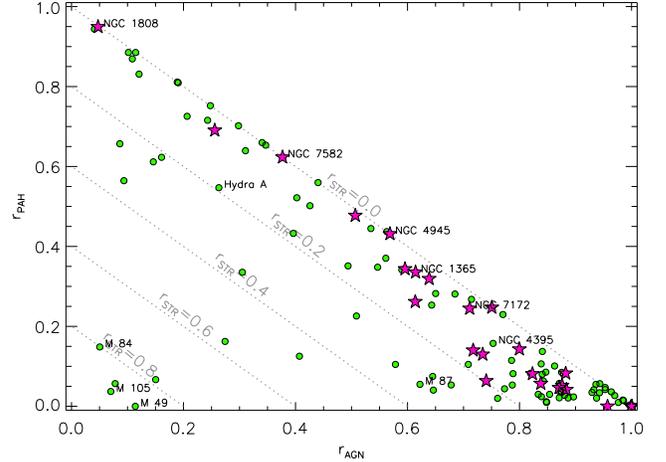}
\caption[]{Comparison of the fractional contributions of the stellar, PAH, and AGN components to the restframe 5--15\um luminosity in the best fitting decomposition model for the IRS spectrum of each source. Big stars represent sources in the subsample with ground-based MIR photometry and spectroscopy, while circles represent those with just photometry.\label{fig:rPAH-rAGN}}
\end{figure}

\begin{figure*} 
\begin{center}
\includegraphics[width=17.7cm]{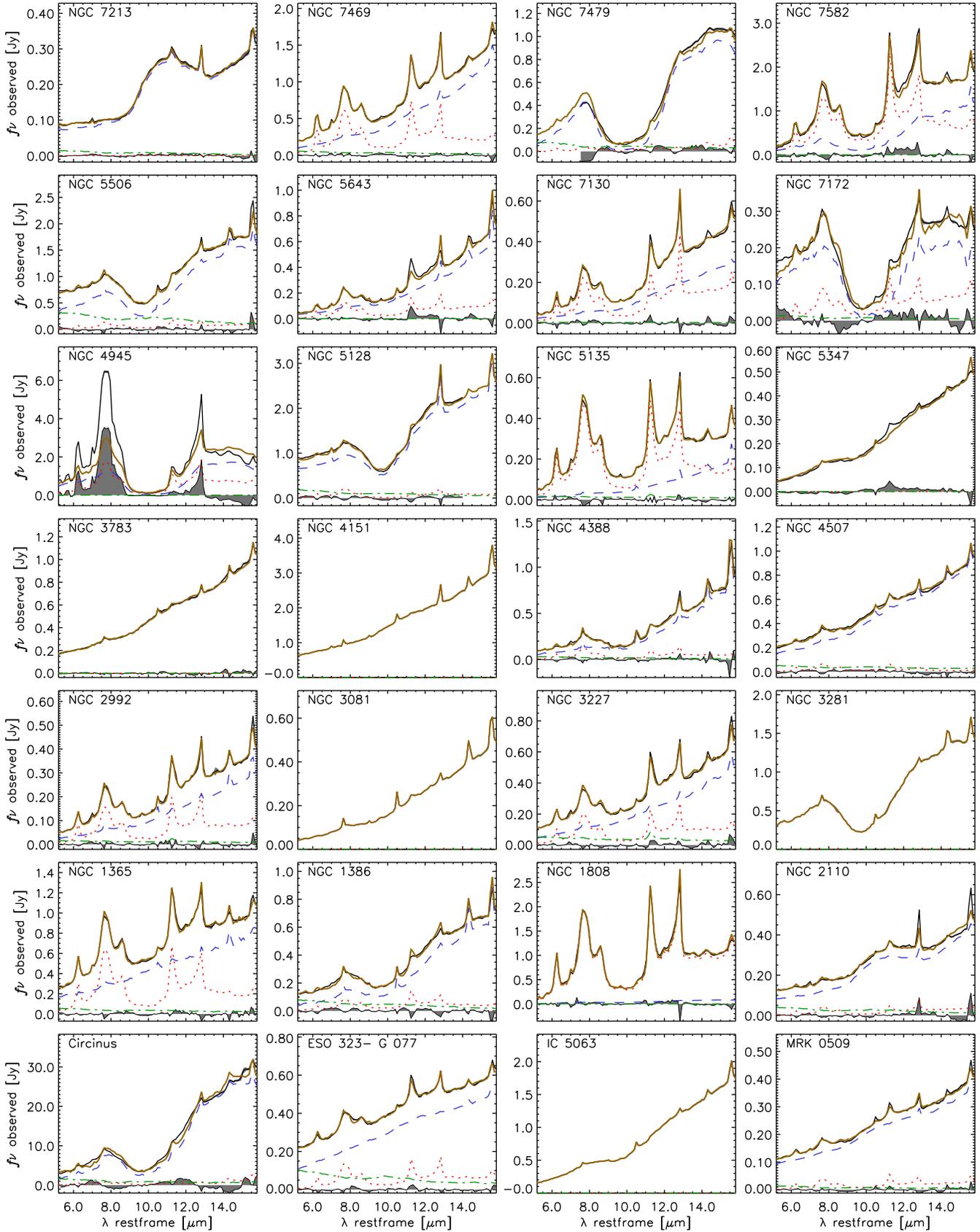}
\end{center}
\caption[]{Best-fit decomposition models for the IRS spectra of the sources with available ground-based MIR spectroscopy. The black solid line represents the original IRS spectrum (resampled at $\Delta\lambda$=0.1\um resolution). The dotted, dashed, and dot-dashed lines represent, respectively, the PAH, AGN, and stellar components of the best-fitting model, shown in yellow. The shaded area at the bottom of each plot represents the residual (spectrum - model). \label{decomposition-plots}}
\end{figure*}

The relative contributions of the stellar, PAH, and AGN components to the luminosity in the 5--15\um range for the best fitting decomposition model (respectively, $r_{\rm{STR}}$, $r_{\rm{PAH}}$, and $r_{\rm{AGN}}$) are shown in Figure \ref{fig:rPAH-rAGN}.
Since by definition these quantities verify $r_{\rm{STR}}$ + $r_{\rm{PAH}}$ + $r_{\rm{AGN}}$ = 1, only two of the three are independent.

Most of the sources, and in particular all the sources in the subsample with ground-based spectroscopy, concentrate between the lines for $r_{\rm{STR}}$ = 0 and 0.2. This reflects the fact that stellar emission usually contributes a very small fraction of the MIR luminosity, except for a few sources with low luminosity AGN and little or no star formation (most of them elliptical galaxies) which lie close to the horizontal axis. However, the blue SED of the stellar templates implies that the stellar components contributes a much larger fraction of the emission at the shortest wavelengths ($\sim$5--6\uu). This allows us to quantify the contribution from stellar emission, which would otherwise degrade the quality of the fits and the derivation of the AGN properties.
The highest concentration of sources in Figure \ref{fig:rPAH-rAGN} is in the region with $r_{\rm{AGN}}$$>$0.7, which is expected given our selection criteria.

Figure \ref{decomposition-plots} shows the spectral decomposition of the IRS spectra for the 28 sources with high-resolution ground-based MIR spectroscopy. 
In most cases the correspondence between the IRS spectrum and the best fitting model is notable. It is particularly notable in sources where the PAH and AGN templates contribute comparable fractions of the MIR luminosity, such as NGC 1365, NGC 2992, or NGC 7582.

The SNR of the IRS spectra is in all cases high enough for the contribution of noise to the residuals to be negligible. Instead, these residuals represent real discrepancies between the spectrum of the source and the decomposition model. While the continuum is usually reproduced with high accuracy, the PAH bands and fine structure lines often show larger residuals. This is because the relative strengths of the individual PAH bands and emission lines vary from source to source, and it is unlikely for any template to match all of them.
In some cases, the best fitting model introduces an excessive amount of PAH component to compensate for an AGN template with too little emission at $\sim$8 and $\sim$12\uu. This is evidenced by negative residuals for most or all of the PAH bands (e.g. NGC 1386, NGC 5347).  
The worst decomposition models are found in some spectra with deep silicate features (NGC 7172, NGC 7479, NGC 4945), which obtain their best fits with OBSCURED templates. This is because most of the IRS spectra with such deep absorptions belong to composite sources which show significant PAH features overlaid on the continuum emission, and therefore they are not eligible as OBSCURED templates. 

\subsection{Measurement of AGN properties}\label{AGNproperties_subsect}

In addition to the fractions contributed by the stellar, PAH, and AGN components to the MIR emission, we use the spectral decomposition to derive other properties of the AGN emission useful for diagnostics. In particular, we obtain expectation values and uncertainties for the monochromatic luminosity of the AGN at 6\um and 12\um restframe (L$^{\rm{AGN}}_{\rm{6}}$, L$^{\rm{AGN}}_{\rm{12}}$), the spectral index of the AGN emission ($\alpha$), and the silicate strength (\ssilu) of the AGN spectrum. 

We obtain the PDF for the fractional contribution of the AGN to the restframe 6\um and 12\um flux densities with the `max' method described in \S\ref{themethod}. Then we multiply this fraction by the restframe 6\um or 12\um monochromatic luminosity of the source in the IRS spectrum (averaged in a 0.2\um wide interval) to obtain L$^{\rm{AGN}}_{\rm{6}}$ and L$^{\rm{AGN}}_{\rm{12}}$, respectively.

To obtain PDFs for $\alpha$ and \ssilu, we need to measure these properties in all the AGN templates first.
We calculate $\alpha$ between the restframe wavelengths 8.1 and 12.5\uu, assuming a power-law continuum of the form $f_\nu \propto \nu^\alpha$. This procedure minimizes the impact of the 10\um silicate feature in the results \citep{Honig10,RamosAlmeida11a}. 

The silicate strength \citep{Levenson07,Spoon07} is defined as:
\begin{equation}
S_{\rm{sil}} = \ln\frac{F(\lambda_p)}{F_C(\lambda_p)}
\end{equation}

\noindent where $F$($\lambda_p$) and $F_C$($\lambda_p$) are the flux densities of the spectrum and the underlying continuum, respectively, at the wavelength of the peak of the silicate feature. We identify this peak by visual inspection \citep[see for details][]{Hatziminaoglou15}. To estimate the continuum underlying the silicate feature we use anchor points at both sides of the silicate feature and interpolate with a power-law.

\begin{figure} 
\begin{center}
\includegraphics[width=8.4cm]{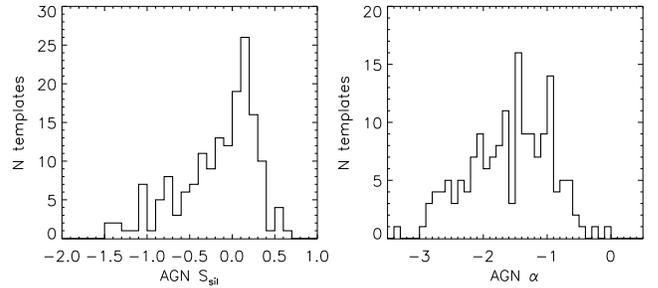}
\end{center}
\caption[]{Distribution of silicate strengths (left) and spectral indices (right) in the library of bona-fide AGN templates.\label{fig:alpha-sil-histogs}}
\end{figure}

As discussed in \S\ref{themethod}, obtaining realistic PDFs for $\alpha$ and \ssil requires that the PDFs to be constructed with a bin width that is large enough to find at least one template in each bin. We use bin widths of 0.1 for both $\alpha$ and \ssilu. These are comparable to the typical uncertainties of $\alpha$ and \ssilu measurements on individual templates, and provide an adequate sampling with no gaps within the normal range of variation of the two quantities (see Figure \ref{fig:alpha-sil-histogs}).

\begin{figure} 
\begin{center}
\includegraphics[width=7.4cm]{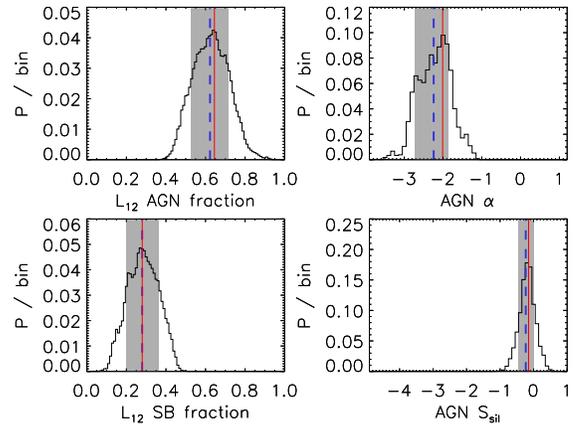}
\end{center}
\caption[]{Examples of probability distributions obtained with the `max' method for several spectral properties of the galaxy NGC 3227. In clockwise order, starting for the top right plot, the represented properties are: the spectral index of the AGN component, the silicate strength of the AGN component, and the fractional contributions from star formation and AGN to the 12\um luminosity within the IRS aperture. In all cases, the histograms represent probability per bin. Bin sizes are 0.01 for L$_{\rm{12}}$ AGN and SB fractions, and 0.1 for $\alpha$ and \ssilu. The dashed line and shaded areas represent, respectively, the expectation value and 1-$\sigma$ confidence interval. The red solid line represents the value derived from the best fitting decomposition model.\label{fig:observables}}
\end{figure}

Figure \ref{fig:observables} shows examples of the PDFs of four spectral properties for one galaxy in the sample, as well as the expectation values and confidence intervals derived from the PDFs. 

Table \ref{table:AGNproperties} lists the expectation values and 1-$\sigma$ confidence intervals (percentiles 16 to 84 of the probability distribution) of L$^{\rm{AGN}}_{\rm{6}}$, L$^{\rm{AGN}}_{\rm{12}}$, $\alpha$, and \ssil for all the sources in the sample. We note that, by construction, the maximum of the PDF corresponds to the value for the best fitting model. Therefore, the expectation value and the value for the best fitting model are usually very close to each other. The only exception is PDFs with multiple peaks or highly asymmetric. In any case, the result for the best fitting model is always within the 1-$\sigma$ confidence interval derived from the PDF.

\begin{figure} 
\includegraphics[width=8.4cm]{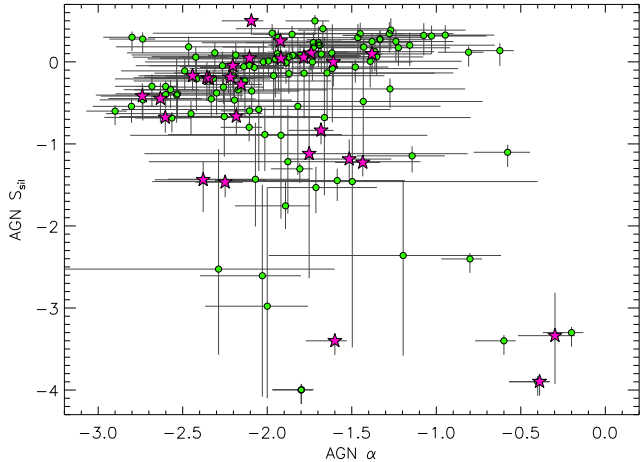}
\caption[]{Expectation values for \ssil and $\alpha$ of the AGN component derived from the full probability distribution functions. Symbols as in Figure \ref{fig:rPAH-rAGN}. Error bars represent the 1-$\sigma$ confidence intervals. Results for sources with $r_{\rm{AGN}}<$0.25 are not shown to improve readability.\label{fig:Ssil-alpha}}
\end{figure}

We compare \ssil and $\alpha$ in Figure \ref{fig:Ssil-alpha}. In sources with \ssilu$>$-1 the best fits are obtained with the bona-fide AGN templates. These sources define a tight sequence where those with \nobreak{$\alpha$$<$-1.7} show increasingly deep silicates for steeper continuum slopes, while for \nobreak{$\alpha$$>$-1.7} the silicate appears mostly in emission and is much less dependent on the spectral index. This is in agreement with results from \citet{Honig10} in a smaller sample. Comparison with predictions from their 3D clumpy torus models (their Figure 12) suggests the relation between $\alpha$ and \ssil is mostly driven by variations in the profile of the radial dust distribution, and consistent with a small dispersion in the average number of clouds along the equatorial line of sight.
On the other hand, sources with deep silicate features (\ssil$<$-1.5) obtain their best fits with some of the OBSCURED templates. They show no obvious relation between \ssil and $\alpha$, as expected if foreground absorption is causing or at least contributing to the depth of the silicate feature \citep[see e.g.][]{Goulding12}. In a large sample of ULIRGs, \citet{Hao07} also find no correlation between $\alpha$ and \ssil when the spectral index is defined between 5 and 15\uu. 
 
\subsection{Comparison with nuclear photometry}

In this section we validate the AGN fractions and AGN luminosities derived from the spectral decomposition model by comparing synthetic photometry on the galaxy-subtracted IRS spectra with the nuclear photometry from high resolution MIR imaging compiled by \citet{Asmus14} in their Table 2.

The sub-arcsecond resolution MIR images were obtained with a variety of instruments and filters in both the N and Q atmospheric transmission bands. The broadband flux of the nuclear source ($f_{\rm{nuc}}$) is measured with procedures that assume either a Gaussian profile with adjustable FWHM (limited to $<$1") or an empirical PSF obtained from the calibration star \citep[see][for the details on the data reduction and photometric measurements.]{Asmus14}.

We select from the entire catalogue of MIR photometric measurements those corresponding to sources with available IRS spectroscopy and obtained with a filter in the N-band.
We also require CV$_{\rm{RMSE}}$$<$0.1 in the best fitting decomposition model (to ignore spurious results from bad fits), and the nuclear source to be detected at the 2$\sigma$ level in either the Gaussian or PSF photometry. There are 456 individual measurements from 113 distinct sources meeting these criteria.

To obtain the synthetic photometry we convolve the IRS SL spectra with the corresponding filter response curve for each of the ground-based measurements. We perform measurements in both the original IRS spectrum ($f_{\rm{tot}}$) and the galaxy-subtracted AGN spectrum ($f_{\rm{AGN}}$). The latter are obtained by subtracting the stellar and PAH components of the best fitting model from the original IRS spectrum.

\citet{Asmus14} found that PSF photometry of the nuclear source is sometimes unreliable due to variation in atmospheric conditions between the observations of the source and the calibration star. They concluded that Gaussian photometry is preferable except for clearly extended nuclei.

We can use the sources that, according to the decomposition, have a very strong AGN component ($r_{\rm{AGN}}$$\ge$0.95) to test the flux cross-calibration of the \textit{Spitzer}/IRS and ground-based observations. Since in these sources the host galaxy contributes $\lesssim$5\% of the flux in the IRS aperture, we expect $f_{\rm{nuc}}$ $\approx$ $f_{\rm{AGN}}$ $\approx$ $f_{\rm{tot}}$.  
We find that Gaussian fluxes are tightly correlated with $f_{\rm{AGN}}$. Their ratios have a symmetric distribution with median 0.998 and 1-$\sigma$ dispersion 0.08 dex or 20\%. This is roughly consistent with estimates of the uncertainty in the absolute flux calibration of 5--10\% for the IRS spectra and 10--15\% for the ground-based photometry.
Ratios calculated from PSF fluxes, on the other hand, show higher dispersion (0.17 dex) and a bias towards $f_{\rm{nuc}}$$<$$f_{\rm{AGN}}$ (median ratio: 1.22). We note that the distribution of ratios for PSF fluxes is also symmetric, suggesting that extended nuclear emission is not the main culprit of their higher dispersion.

\begin{figure} 
\includegraphics[width=8.4cm]{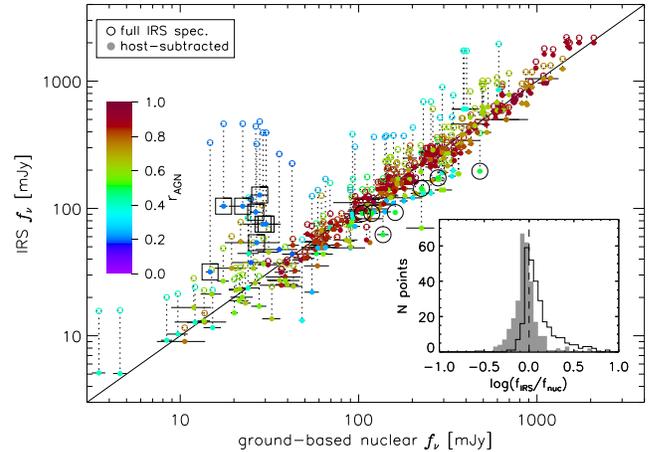}
\caption[]{Comparison between MIR fluxes from synthetic photometry extracted from the IRS spectra ($f_{\rm{IRS}}$) and nuclear fluxes obtained extracted with the Gaussian profile ($f_{\rm{nuc}}$) for sources with $r_{\rm{AGN}}$$<$0.95. IRS fluxes are extracted for both the integrated host+AGN spectrum (open symbols) and the AGN component (solid symbols). Vertical dotted lines connect measurements corresponding to the same source and filter. The diagonal line marks the 1:1 relation. Big open squares and circles mark measurements for NGC 1097 and NGC 7130, respectively (see text). The color coding indicates the fractional contribution from the AGN to the 5--15\um luminosity derived from the spectral decomposition. Horizontal bars indicate the photometric errors of nuclear measurements. The inset plot shows the distribution of the ratios between AGN (filled histogram) or total IRS (open histogram) and Gaussian nuclear fluxes.
\label{fig:compare-fluxes2}}
\end{figure}

Figure \ref{fig:compare-fluxes2} compares $f_{\rm{tot}}$ and $f_{\rm{AGN}}$ from the IRS spectrum with Gaussian nuclear fluxes from the ground-based photometry for the remainder of the sample, that is, the sources with $r_{\rm{AGN}}$$<$0.95. Dotted lines connect values of $f_{\rm{tot}}$ (open symbols) and $f_{\rm{AGN}}$ (solid symbols) corresponding to the same nuclear photometric measurement, and represent the decrease in the flux density caused by the subtraction of the stellar and PAH components from the IRS spectrum. We have color-coded the symbols according to the $r_{\rm{AGN}}$ of the sources.

The correlation between total IRS fluxes and nuclear fluxes is not as tight as in the $r_{\rm{AGN}}$$\ge$0.95 subsample, and there are important deviations (always in the sense $f_{\rm{tot}}$$>$$f_{\rm{nuc}}$) in many of the sources with $r_{\rm{AGN}}$$<$0.5. The distribution of $f_{\rm{tot}}$/$f_{\rm{nuc}}$ (empty histogram in Figure \ref{fig:compare-fluxes2}) peaks at 1:1 but is highly asymmetric, with a long tail for ratios $>$1. This is the expected result, since host emission arises mostly from outside the region sampled by nuclear photometry. 
However, when we consider the host-subtracted AGN spectra, the tight correlation with the nuclear fluxes is restored, and it holds remarkably well even for sources with AGN fractions as low as $r_{\rm{AGN}}$$\sim$0.3 (cyan symbols).

\begin{figure}
\begin{center}
\includegraphics[width=8.4cm]{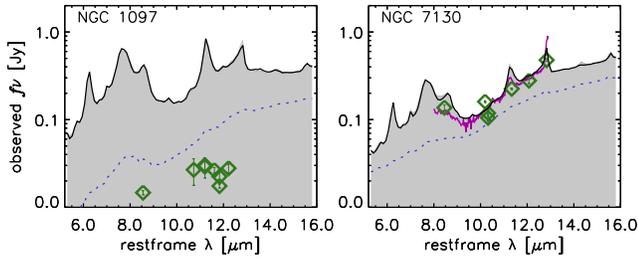}
\end{center}
\caption[]{Comparison of AGN components derived from the best fitting decomposition model with nuclear photometry and spectroscopy from ground-based observations for the outliers in the relation shown in Figure \ref{fig:compare-fluxes2}: NGC 1097 (left) and NGC 7130 (right). The solid area represents the observed IRS spectrum and the solid line its best fitting decomposition model. The dotted line represents the AGN component. Diamonds with error bars represent the nuclear photometry. The magenta line represents the nuclear spectrum of NGC 7130.\label{fig:NGC1097}}
\end{figure}

The distribution of $f_{\rm{AGN}}$/$f_{\rm{nuc}}$ in the $r_{\rm{AGN}}$$<$0.95 subsample (filled histogram in the inset plot in Figure \ref{fig:compare-fluxes2}) is not completely symmetric, and it is slightly biased towards $f_{\rm{AGN}}$ $<$ $f_{\rm{nuc}}$ (median value: 0.895). This could indicate a systematic underestimation of $\sim$10\% in the AGN fluxes from the spectral decomposition. However, it could also indicate that some of the nuclear emission (10\% on average) arises in the host galaxy, not the AGN. The detection of PAH features in the sub-arcsecond resolution nuclear spectra of some of these sources favours the latter (see \S\ref{comp-ground-spec}).
The boldest such case, marked with big open circles in Figure \ref{fig:compare-fluxes2}, is NGC 7130. This galaxy hosts a Compton thick AGN and a nuclear starburst \citep{Levenson05,Gonzalez-Delgado98} that is known to contribute a significant fraction of the MIR emission even on sub-arcsecond scales \citep{Alonso-Herrero06,Diaz-Santos10}. Interestingly, the spectral decomposition finds that the AGN contributes $\sim$60\% of the nuclear luminosity in the 8--12\um range (Figure \ref{fig:NGC1097}, right panel).

Nearly all the highest values of $f_{\rm{AGN}}$/$f_{\rm{nuc}}$ correspond to measurements from a single galaxy, NGC 1097 (marked with big open squares in the figure). This galaxy has a low luminosity LINER or Seyfert 1 nucleus as well as a young, compact ($<$9pc) nuclear star cluster \citep{Mason07}. The IRS spectrum from CASSIS also includes extended emission from a kpc-scale circumnuclear star formation ring.
It is striking that the best decomposition model fits remarkably well the IRS spectrum (Figure \ref{fig:NGC1097}, left panel), yet it predicts AGN fluxes 2--5 times higher than observed for the nuclear source, which could itself be dominated by emission from dust heated by the star cluster instead of the AGN \citep{Mason07}.
This source has one of the lowest AGN fractions in the sample ($r_{\rm{AGN}}$=0.20). At such low fractions, the relative errors in the predicted AGN fluxes are large.
While the 1-$\sigma$ confidence interval for the 12\um flux density of the AGN is 60--130 mJy (compared to $\sim$25 mJy for the nuclear source), the PDF is wide and it estimates at 5\% the probability of $f_{\rm{AGN}}$$<$25 mJy.

Excluding these two outliers, the 1-$\sigma$ dispersion in $f_{\rm{AGN}}$/$f_{\rm{nuc}}$ for the galaxies with $r_{\rm{AGN}}$$<$0.95 is 0.12 dex or 32\%. Since the dispersion in the $r_{\rm{AGN}}$$\ge$0.95 subsample is 0.08 dex (20\%) due to photometric uncertainties alone, we estimate a 25\% uncertainty in the fractional contribution of the AGN to the observed IRS fluxes. This is in good agreement with a mean value of 23\% for the relative error in the 12\um AGN luminosities derived from the decomposition, and confirms that PDFs derived with the `max' method provide reliable uncertainties.

\subsection{Comparison with ground-based sub-arcsecond MIR spectra\label{comp-ground-spec}}

\begin{figure*}
\begin{center}
\includegraphics[width=17.7cm]{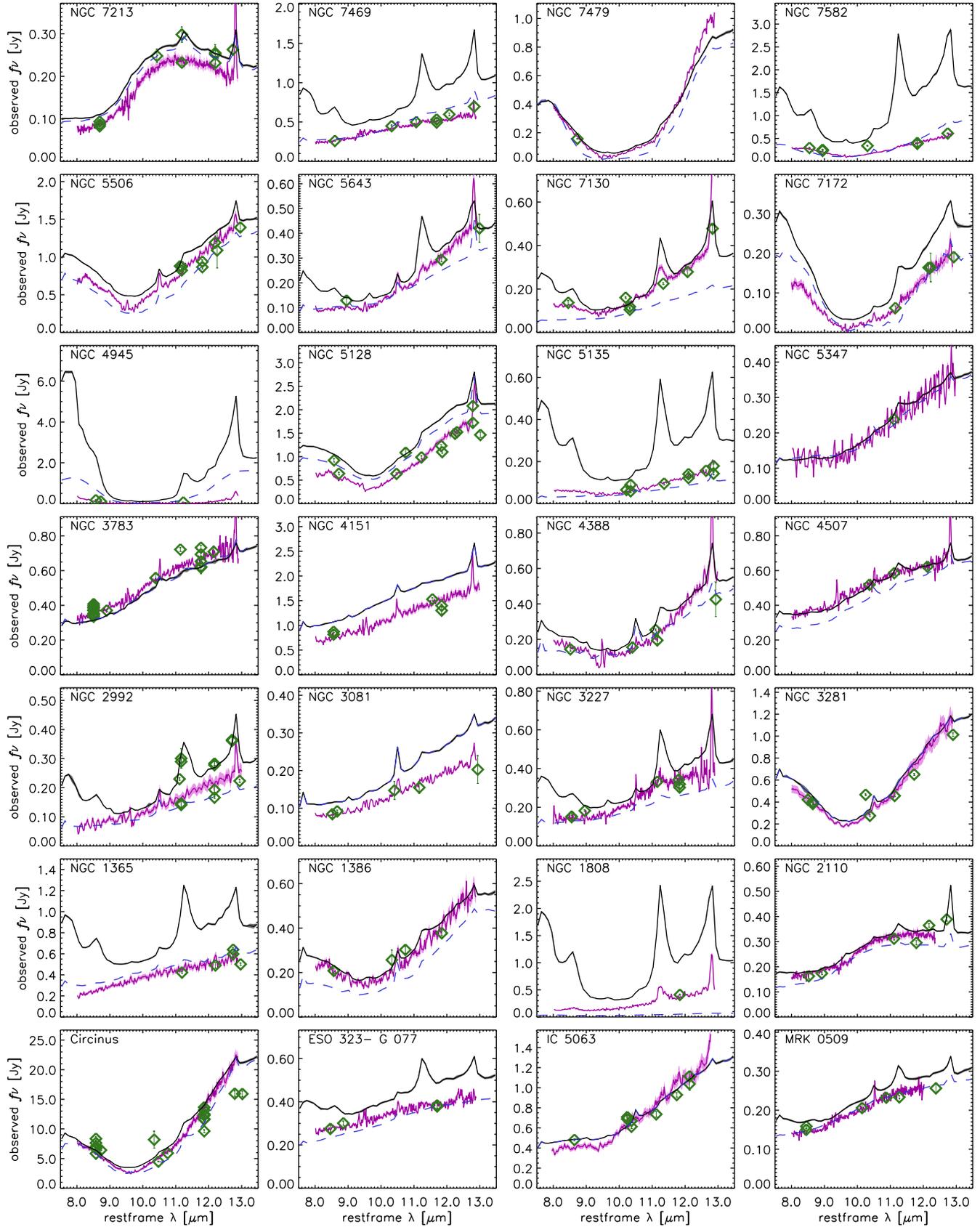}
\end{center}
\caption[]{Comparison of IRS decomposition results with the ground-based photometry and spectroscopy. The solid black and blue dashed lines represent the restframe 7.5--13.5\um section of the IRS spectrum and the AGN component, respectively. The green diamonds represent the Gaussian fluxes for the nuclear point source in the catalogue from \citet{Asmus14}, with error bars representing 1-$\sigma$ uncertainties. The magenta solid line represents the full-resolution ground-based spectrum scaled to fit the Gaussian photometry. The light magenta shaded area represents the 1-$\sigma$ uncertainty in the scaling factor for the nuclear spectrum.\label{compare-spectra}}
\end{figure*}

There are 28 sources where we can directly compare the galaxy-subtracted AGN spectrum from the IRS spectroscopy with a sub-arcsecond resolution ground-based spectrum. 

The absolute flux calibration of the ground-based nuclear spectra has larger uncertainties compared to both the photometry from ground-based imaging and IRS spectra mostly due to slit losses. To overcome this, we have re-scaled the nuclear spectra to match the available nuclear photometry. 
Scaling factors are calculated as the ratio between fluxes from the nuclear (Gaussian) photometry and synthetic fluxes measured on the nuclear spectrum with the same filters.
Since most sources have several photometric measurements, we obtain a weighted average of all the scaling factors for a given source, with weights: $w_i$ = 1/$\sigma_i^2$, where $\sigma_i$ is the uncertainty in the scaling factor for the $i^{th}$ measurement.

Figure \ref{compare-spectra} compares the AGN spectra derived from the decomposition with the nuclear spectroscopy and photometry from the ground. With few exceptions, there is a remarkable correspondence between the \textit{shape} of the observed nuclear spectra and those predicted by the decomposition model. This correspondence is particularly interesting in sources with a strong PAH component in the IRS spectra, such as NGC1365, NGC2992, NGC 3227, or NGC 7469, where it would be very difficult to constrain the AGN luminosity (let alone its SED) by other means.
There are, however, significant discrepancies in the normalization of many sources, with nuclear fluxes up to $\sim$30\% lower or higher compared to IRS AGN fluxes. 
While uncertainty in the absolute calibration of both the IRS spectra and the nuclear photometry contribute to these discrepancies, it is insufficient to explain the extreme cases. In the following we discuss other interpretations for the origin of these discrepancies that apply to subgroups of sources.

Some sources with ground-based nuclear fluxes significantly ($\gtrsim$20\%) higher than IRS AGN fluxes have clear PAH emission features in their nuclear spectra that evidence nuclear starbursts. Well known cases are NGC 1808 \citep[e.g.][]{Tacconi-Garman96,Sales13}, NGC 3227 \citep{Davies06,Davies07,Honig10}, and NGC 7130 \citep{Diaz-Santos10}. 
In these sources, the nuclear starburst contributes a fraction of the unresolved nuclear MIR emission, which is therefore higher than the emission from the AGN alone. Accordingly, AGN fluxes derived from the decomposition might be better estimates of their actual AGN output. On the other hand, if no high angular-resolution spectrum is available, the spectral decomposition does not provide any information as to whether there is nuclear star formation or not.

Other sources show a slight nuclear excess compared to the IRS AGN flux in spite of very weak or absent PAH features in the nuclear spectrum (NGC 1386, NGC 4507). An inspection of the spectral components and residuals in Figure \ref{decomposition-plots} reveals that the AGN component derived from the decomposition is probably underestimated in these two sources. Their residuals resemble an inverted PAH spectrum, indicating the PAH component has been boosted at the expense of the AGN to improve the fit of a pronounced 8\um bump.

In NGC 3783 the nuclear spectrum is slightly above the original IRS spectrum. This is unphysical, since the IRS aperture is much wider than the ground-based one. Inspection of the pointing information in the IRS spectrum reveals that the nuclear source was slightly off-center in the IRS slit.

On the other hand, the cases where the AGN component in the IRS spectrum is above the nuclear spectrum could be explained by dust heated by the AGN on arcsecond scales (i.e. dusty ionisation cones). The Gaussian photometry from \citet{Asmus14} does not recover all the nuclear flux if the source has a FWHM$>$1'' or the profile is less concentrated than a Gaussian. This seems to be the case at least for NGC 3081, NGC 4151, and NGC 5128, where Gaussian fluxes are higher than PSF fluxes, yet lower than IRS AGN fluxes.

While uncertainties in the absolute flux calibrations and the physical extension of the AGN-heated MIR emitting region introduce significant dispersion between ground-based and IRS fluxes, the \textit{shape} of the AGN SED should be relatively unaffected and therefore provides a more demanding test of the decomposition model. 

To quantify the differences in the SED of the nuclear point source (as represented by the ground-based spectrum) and the AGN spectrum derived from the decomposition model, we use the spectral index of the continuum ($\alpha$) and the strength of the 10\um silicate feature (\ssilu) as indicators. 

We calculate $\alpha$ for the nuclear spectra using the same method applied to the AGN templates (see \S\ref{AGNproperties_subsect}), with anchor points at 8.1 and 12.5\uu.
For the two sources whose nuclear spectra are truncated shortwards of 12.5\uu, we use 8.1 and 12.0\um instead. We estimate the uncertainty in $\alpha$ via error propagation assuming a 5\% uncertainty in the fluxes. 

To estimate \ssil in the nuclear spectra we place the first anchor point for the underlying continuum at 8.1\um and the second somewhere between 12.7 and 13\uu, putting great care in avoiding noisy regions and the 12.81\um \neii line.  As we did for the AGN templates, we identify the peak wavelength of the silicate feature by visual inspection.
Error bars for \ssil are estimated by performing Monte Carlo simulations in which flux densities are varied within their uncertainties. We note that these errors account for the uncertainty in \ssil due to noise in the spectra, but do not include the systematic error introduced by the selection of the continuum anchor points and the uncertainty in the wavelength of the peak of the silicate feature.

\begin{figure} 
\includegraphics[width=8.4cm]{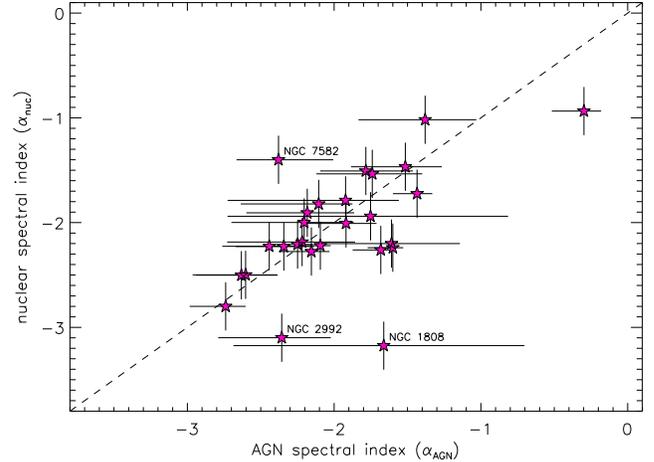}
\caption[]{Comparison between spectral indices measured directly on the nuclear spectra and those for the AGN component derived from the spectral decomposition of the IRS spectrum. The dashed line marks the 1:1 relation.\label{fig:alpha-comp}}
\end{figure}

Figure \ref{fig:alpha-comp} shows the spectral index of the ground-based nuclear spectrum ($\alpha_{\rm{nuc}}$) versus the expectation value for the AGN component in the IRS spectrum ($\alpha_{\rm{AGN}}$). There is a good correlation between the two, albeit with a significant dispersion. The largest discrepancies correspond to sources with very weak AGN components (NGC 1808, NGC 7582) or poorly fit decomposition models (NGC 4945, NGC 7172). For the remaining sources $\alpha_{\rm{nuc}}$ and $\alpha_{\rm{AGN}}$ values are compatible within their uncertainties. The median value of the difference in spectral indices, $\vert \alpha_{\rm{AGN}} - \alpha_{\rm{nuc}} \vert$, is 0.19.
 
\begin{figure} 
\includegraphics[width=8.4cm]{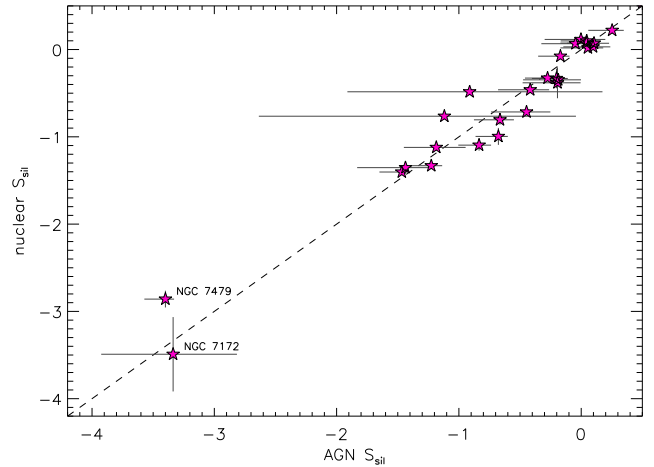}
\caption[]{Comparison between silicate strengths (\ssilu) measured directly on the nuclear spectra and those for the AGN template in the best fitting decomposition model for the IRS spectrum.\label{fig:Ssil-comp}}
\end{figure}

Figure \ref{fig:Ssil-comp} compares the \ssil from the nuclear spectra and the AGN component in the IRS spectra. There is a good agreement between the two, even in the sources with the deepest silicates (NGC 4945, NGC 7172) which obtain poor fits with large residuals. The median absolute difference in \ssil measurements is 0.10, with a small bias towards smaller nuclear \ssil values compared to those from the AGN component on the IRS spectra.

Obtaining such tight correlations for both $\alpha$ and \ssil confirms that the spectral decomposition successfully separates the AGN and host galaxy components of the emission, and that the AGN properties derived are reliable (within their uncertainties). We emphasize that even at the sub-arcsecond resolution of ground-based spectroscopy, nuclear spectra do not always represent accurately the AGN emission, which can be extended or overlap with a nuclear starburst. This is, arguably, one of the factors contributing to the scatter in Figures \ref{fig:alpha-comp} and \ref{fig:Ssil-comp}, which would be even smaller if a better proxy for the actual AGN spectrum could be found.

\section{Discussion}

We have found that the main limitation of our decomposition method is the existence of categories of sources whose spectra are not well reproduced, at least with the current library of templates. We have identified galaxies with deep silicate absorptions as one important such category. We consider that the reason for the poor fits in galaxies with deep absorptions is the small number of suitable templates in the library, which is caused by the difficulty in finding pure-AGN and pure-PAH spectra with deep absorption features within the limited (albeit large) set of available IRS observations, since many of the spectra with such deep features belong to composite sources.
The relatively small size and restricted selection criteria of our test sample implies that we cannot rule out other categories of sources which could also prove difficult to model with the current library. Even for source types that typically obtain good fits, there probably exist extreme cases of exotic spectra whose properties are beyond the normal range of variation covered by the templates. However, we expect these to be few in numbers.

Degeneracy was also a legitimate concern, since the large number of available templates suggested there could be many combinations of them capable of producing good fits with comparable $\chi^2$ values for any particular source. However, we found that while this is typically the case, the fractional contribution to the MIR emission from each of the three spectral components is usually well constrained. The reason for this remarkable lack of degeneracy is probably the singular characteristics of the 5--15\um spectral range, where the spectral shapes of the stellar, interstellar (PAH), and AGN emissions are clearly different from each other.

Because the spectra for the stellar and PAH components show very limited variation, we can recover not just the strength of the AGN emission but also its spectral shape. Decomposition models that obtain good fits for a particular source usually involve AGN templates very similar to each other, indicating that it is not easy for the stellar and PAH templates to compensate for an AGN template with, for example, the wrong spectral index or silicate strength. 

The presence in the library of many similar (but not identical) templates is not unnecessary redundancy: they help improve the fits by increasing the chances of reproducing fine details in the spectrum, like emission lines. This is important because in high SNR spectra these fine details dominate over photometric errors in the residuals and cause large $\chi^2$ values if the template set is not varied enough. 
Having multiple similar templates does not complicate the calculation of PDFs for AGN properties thanks to the mechanics of the `max' method. In fact, the resolution of the PDFs is limited by the size of the template set, since at least one template in each bin is required. More detailed PDFs allow for better estimates of the uncertainties in AGN properties.

We have compared our results from the spectral decomposition with ground-based high spatial resolution imaging and spectroscopy as the ultimate test for our method. There is a general agreement between galaxy-subtracted AGN and nuclear fluxes within their uncertainties, and a good correspondence in $\alpha$ and \ssilu. This has positive implications for the analysis with this method of the IRS spectra of AGN at higher redshifts \citep[see][]{Hatziminaoglou15}, or those to be obtained in the near future with the Mid-InfraRed Instrument (MIRI) onboard the James Webb Space Telescope (JWST).
However, there are also discrepancies, which in some cases suggest a poor decomposition and in others seem caused by a significant contribution from the host even at the sub-arcsecond scales of the nuclear photometry and spectroscopy.  

The sensitivity of the fit to the shape of the AGN template depends on the relative contribution of the AGN component to the MIR emission within the IRS aperture. When it is low, the uncertainty in the spectral shape of the AGN is large. This means that AGN properties other than its luminosity are poorly constrained if the host galaxy strongly dominates the emission within the extraction aperture. This fundamental limitation of the decomposition method implies that high spatial resolution observations are necessary if the AGN does not dominate the integrated emission of the galaxy. Furthermore, even for AGN-dominated sources, ground-based high angular resolution observations are still required to infer the spatial distribution of the host emission (for example, to distinguish between nuclear, circumnuclear, or galaxy-wide star formation) and to detect weak fine structure lines.

\section{Summary and Conclusions}\label{sect-summary}

In this paper we have demonstrated a new data-driven method for spectral decomposition of mid-infrared spectra that uses as templates sets of real spectra from individual sources whose emission is strongly dominated by a single physical component.
We have shown that this decomposition method produces fits with an excellent accuracy for the \textit{Spitzer}/IRS spectra of composite sources, with residuals a factor $\sim$2 lower than those of a competing algorithm with a higher number of free parameters. While some peculiar spectra obtain poor fits due to their physical properties falling outside the range of variation available from our template set, the combined $\chi^2_\nu$ and CV$_{\rm{RMSE}}$ criteria allow us to identify these cases and set them apart for a customized analysis.

We have applied the `max' method to obtain PDFs for model parameters and physical properties of the AGN. Taking into account the full PDF instead of just the best fitting model enables an adequate treatment of degeneracy in the parameters derived from the decomposition models, which allows for realistic uncertainties.
We have obtained restframe 6\um and 12\um luminosities, spectral indices ($\alpha$), and silicate strengths (\ssilu) for the AGN spectra after removal of the emission from the host. We find no clear trend of either $\alpha$ or \ssil with the AGN luminosity. There is a tight correlation between $\alpha$ and \ssil that applies to all AGN except those with deep silicate absorption, where strong reprocessing by dust erases the signature of the dominant power source.

To validate the AGN luminosities derived from the spectral decomposition we have compared sub-arcsecond resolution N-band photometry from ground-based observations with synthetic photometry derived from the IRS spectra before and after subtraction of the host emission.
We find a good agrement between nuclear fluxes and those from the AGN component in the IRS spectrum, with a typical dispersion of 0.12 dex in the ratio between the two. We conclude that this dispersion is dominated by uncertainty in the relative flux calibration between the ground-based photometry and the IRS spectra.

We have further validated the decompositions using a subsample of 28 sources with ground-based high angular resolution 8--13\um spectra. We find a remarkable correspondence between the shape (as described by the spectral index $\alpha$ and the silicate strength \ssilu) of the observed nuclear spectra and the AGN spectrum derived from the decomposition model, even in sources whose IRS spectrum is dominated by emission from the host galaxy. We interpret discrepant cases as caused by either extended AGN emission or nuclear starbursts.
The decomposition of IRS spectra predicts $\alpha$ and \ssil of the nuclear spectra with a typical accuracy of 0.19 and 0.10, respectively. These small errors confirm that our spectral decomposition is successful at separating the AGN and host component of the emission and provides a reliable estimate of the shape of the AGN spectrum. 
This allows for unbiased studies of the AGN emission in intermediate and high redshift galaxies, which are inaccesible to current ground-based instruments, using archival \textit{Spitzer}/IRS observations and in the near future with JWST/MIRI.
The decomposition code and templates are available at \url{http://www.denebola.org/ahc/deblendIRS}.

\acknowledgements
We thank the anonymous referee for their useful comments that helped to improve this paper. A.H.-C. and A.A.-H. acknowledge funding by the Universidad de Cantabria Augusto Gonz\'alez Linares programme and the Spanish Plan Nacional de Astronom\'ia y Astrof\'isica under grant AYA2012-31447. 
C.R.A. is supported by a Marie Curie Intra European Fellowship within the 7th European Community Framework Programme (PIEF-GA-2012-327934). S.F.H. acknowledges support from the Marie Curie International Incoming Fellowship within the 7th European Community Framework Programme (PIIF-GA-2013-623804). T.D.S. is supported by ALMA-CONICYT grant number 31130005. P.E. acknowledges support from the Spanish Programa Nacional de Astronom\'{\i}a y Astrof\'{\i}sica under grant AYA2012-31277. 
This work is based on observations made with the \textit{Spitzer Space
Telescope}, which is operated by the Jet Propulsion Laboratory, Caltech
under NASA contract 1407.
The Cornell Atlas of Spitzer/IRS Sources (CASSIS) is a product of the Infrared Science Center at Cornell University, supported by NASA and JPL.

\begin{appendix}

\section{Comparison with DecompIR}

To asses the accuracy of the fits obtained with our decomposition method, we compared the residuals of the best fitting model produced by our method with those from another decomposition routine, DecompIR \citep{Mullaney11}. We chose DecompIR because it produces very good fits for a wide variety of \textit{Spitzer}/IRS spectra and is frequently used in the recent literature \citep[e.g.][]{Goulding12,Magdis13,Drouart14}.

For the comparison we used the \textit{Spitzer}/IRS spectra of the 118 sources in our samples that were not selected as AGN templates. 
We pre-processed the spectra using the exact same procedures for de-redshifting, rebinning, and truncation of the $\lambda$$>$15.8\um portion of the spectra. Because the starburst templates from DecompIR are not defined for $\lambda$$<$6\uu, we also truncated this section of the spectra.
We used the latest public version of DecompIR\footnote{Version 1.21; available at https://sites.google.com/site/decompir/}, with all nine parameters defining the model set free. We ran DecompIR five times for every source, each time with one of the five different starburst templates, and took for each source the solution with the lowest $\chi^2$.

\begin{figure} 
\includegraphics[width=8.4cm]{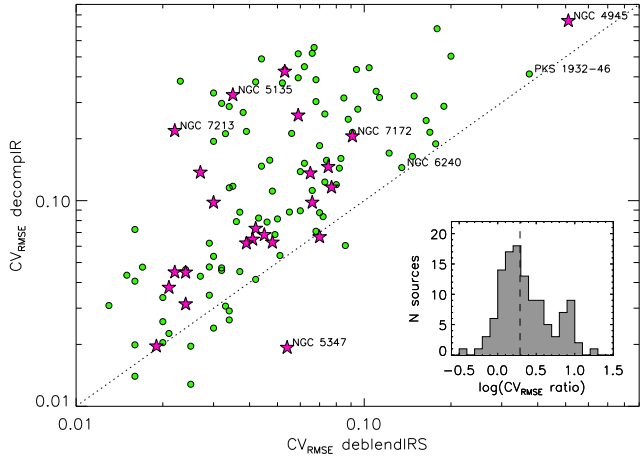}
\caption[]{Comparison of the coefficient of variation of the root-mean-square error (CV$_{\rm{RMSE}}$) for the best fitting model obtained with DecompIR and our decomposition routine. Symbols as in Figure \ref{fig:rPAH-rAGN}. The dotted line represents the 1:1 relation. The inset plot represents the distribution of the ratio between CV$_{\rm{RMSE}}$ values obtained with the two routines (solid histogram) and its median value (dashed line).\label{fig:residuals}}
\end{figure}

For each source, we computed the coefficient of variation of the root-mean-square error (CV$_{\rm{RMSE}}$; see Eq. 3) for the best fitting decomposition model obtained with our method and DecompIR. Figure \ref{fig:residuals} compares the two values for the 118 sources. It shows that residuals from DecompIR are usually larger, with 90\% of sources lying above the line that marks the 1:1 relation. The median ratio of the CV$_{\rm{RMSE}}$ is 1.95, which indicates that the typical residuals with our routine are a factor $\sim$2 smaller than those obtained with DecompIR. In some extreme cases, residuals can be up to a factor $\sim$10 smaller (sources in the top left corner of Figure \ref{fig:residuals}).
While DecompIR produces smaller residuals in 10\% of the sources, nearly all of these obtain very good fits with both methods (CV$_{\rm{RMSE}}$$<$0.04).

Interestingly, all the sources with deep silicate features that obtained poor fits (large residuals) with our routine also obtain equally large or even larger residuals with DecompIR. This is despite the fact that DecompIR uses two free parameters to fit the extinction in the AGN and the host separately. By default, DecompIR limits the range of available optical depths at 9.7\uu, $\tau_{\rm{9.7}}$, to the interval [0,5] for both the AGN and the host. In many of the sources with deep silicates the best fitting model uses $\tau_{\rm{9.7}}$=5 for either the AGN or the host (or both), suggesting a wider range is needed. After increasing the maximum $\tau_{\rm{9.7}}$ to 15, some of the sources obtain marginally lower $\chi^2$ with higher extinction values. However, the impact on the residuals is negligible.

\begin{figure} 
\includegraphics[width=8.4cm]{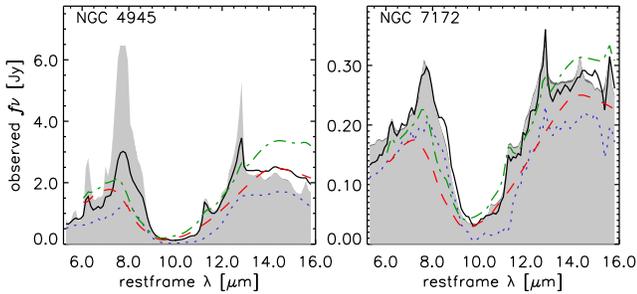}
\caption[]{Comparison of decomposition models for two sources with deep silicate features. The shaded area represents the rebinned IRS spectrum. The best fitting model for the combined host+AGN emission is shown in solid line (for DeblendIRS) and dot-dashed line (for DecompIR). The AGN spectra fitted by DeblendIRS and DecompIR are represented by the dotted and dashed lines, respectively.\label{fig:fitcompdeepsil}}
\end{figure}

In Figure \ref{fig:fitcompdeepsil} we show as examples the best fitting models from DeblendIRS and DecompIR for the two sources in the subsample with ground-based spectroscopy that obtained the worst residuals with DeblendIRS: NGC 4945 and NGC 7172.
Residuals with DecompIR are even larger, indicating that this kind of sources are difficult to fit also for methods that use a freely varying extinction coefficient.

\end{appendix}

\clearpage

\begin{deluxetable*}{lrlrlrlr r r}
\centering
\tabletypesize{\scriptsize}
\tablewidth{0pc}
\tablecolumns{10}
\tablecaption{Parameters for best-fitting decomposition model\label{table:decomposition}}
\tablehead{\colhead{Source name} & \colhead{$z$} & \colhead{PAH template} & \colhead{r$_{PAH}$} & \colhead{AGN template} & \colhead{r$_{AGN}$} & \colhead{stellar template} & \colhead{r$_{STR}$} & \colhead{$\chi^2_{\nu}$} & \colhead{CV$_{RMSE}$}}
\startdata
ARK 120                   &    0.0327 & MCG -02-33-098            &    0.036 & [HB89] 1435-067           &    0.875 & NGC 5831                  &    0.089 &      0.55 &    0.016 \\
CGCG 041-020              &    0.0360 & 2MASX J14311978+3534184   &    0.342 & ESO 103- G 035            &    0.590 & NGC 4377                  &    0.068 &      1.41 &    0.034 \\
Circinus                  &    0.0014 & NGC 3187                  &    0.046 & IRAS  12071-0444          &    0.871 & IC 2006                   &    0.083 &     13.45 &    0.070 \\
ESO 005- G 004            &    0.0062 & ESO 353- G 020            &    0.268 & NGC 6300                  &    0.714 & NGC 5831                  &    0.018 &      6.80 &    0.060 \\
ESO 121-IG 028            &    0.0405 & NGC 3187                  &    0.044 & ESO 103- G 035            &    0.773 & MESSIER 085               &    0.183 &      0.63 &    0.055 \\
ESO 138- G 001            &    0.0091 & ARP 256 NED01             &    0.055 & VII Zw 244                &    0.936 & NGC 5831                  &    0.009 &      2.14 &    0.034 \\
ESO 140- G 043            &    0.0142 & NGC 3310                  &    0.101 & PG 1302-102               &    0.863 & NGC 5812                  &    0.037 &      0.49 &    0.016 \\
ESO 141- G 055            &    0.0371 & NGC 3801                  &    0.023 & 3C 445                    &    0.896 & NGC 1700                  &    0.080 &      0.23 &    0.013 \\
ESO 198- G 024            &    0.0455 & NGC 3310                  &    0.035 & PG 1302-102               &    0.929 & NGC 4660                  &    0.036 &      0.49 &    0.027 \\
ESO 286-IG 019            &    0.0430 & MCG +08-11-002            &    0.281 & NGC 4418                  &    0.685 & NGC 5831                  &    0.034 &      7.50 &    0.073 \\
ESO 297- G 018            &    0.0252 & NGC 0877                  &    0.253 & IC 5063                   &    0.643 & NGC 1700                  &    0.104 &      1.67 &    0.037 \\
\enddata
\tablecomments{Table 1 is published in its entirety in the electronic edition of the Astrophysical Journal. A portion is shown here for guidance regarding its form and content.}
\end{deluxetable*}

\begin{deluxetable*}{lr c c c c c c c c}
\tabletypesize{\scriptsize}
\tablewidth{0pc}
\tablecolumns{10}

\tablecaption{AGN properties derived from decomposition\label{table:AGNproperties}}
\tablehead{\colhead{Source name} & \colhead{$z$} & \colhead{log L$_{\rm{6}}^{\rm{AGN}}$/erg s$^{-1}$} & \colhead{1-$\sigma$ interval} & \colhead{log L$_{\rm{12}}^{\rm{AGN}}$/erg s$^{-1}$} & \colhead{1-$\sigma$ interval} &\colhead{$\alpha$ AGN} & \colhead{1-$\sigma$ interval} & \colhead{\ssil AGN} & \colhead{1-$\sigma$ interval}}
\startdata
ARK 120                   &    0.0327 &   43.98 &     43.80 -- 44.09 &   44.03 &     43.96 -- 44.07 &   -1.03 &     -1.45 -- -0.70 &    0.31 &      0.09 --  0.44 \\
CGCG 041-020              &    0.0360 &   42.75 &     42.53 -- 42.89 &   43.16 &     43.06 -- 43.23 &   -2.01 &     -2.59 -- -1.56 &   -0.89 &     -1.33 -- -0.56 \\
Circinus                  &    0.0014 &   42.52 &     42.44 -- 42.66 &   43.11 &     43.10 -- 43.13 &   -2.25 &     -2.42 -- -2.14 &   -1.47 &     -1.65 -- -1.40 \\
ESO 005- G 004            &    0.0062 &   41.80 &     41.72 -- 41.85 &   42.06 &     42.01 -- 42.10 &   -1.71 &     -2.00 -- -1.35 &   -1.53 &     -1.85 -- -1.28 \\
ESO 121-IG 028            &    0.0405 &   42.46 &     42.15 -- 42.66 &   42.93 &     42.85 -- 42.99 &   -2.25 &     -2.87 -- -1.69 &   -0.67 &     -1.15 -- -0.26 \\
ESO 138- G 001            &    0.0091 &   43.22 &     43.11 -- 43.31 &   43.47 &     43.45 -- 43.49 &   -1.74 &     -2.00 -- -1.52 &    0.12 &     -0.09 --  0.23 \\
ESO 140- G 043            &    0.0142 &   43.32 &     43.24 -- 43.40 &   43.52 &     43.48 -- 43.55 &   -1.70 &     -2.00 -- -1.42 &    0.21 &      0.01 --  0.30 \\
ESO 141- G 055            &    0.0371 &   43.91 &     43.75 -- 44.00 &   44.03 &     43.98 -- 44.06 &   -1.33 &     -1.77 -- -1.03 &    0.27 &      0.05 --  0.38 \\
ESO 198- G 024            &    0.0455 &   43.36 &     43.21 -- 43.45 &   43.58 &     43.53 -- 43.61 &   -1.73 &     -2.12 -- -1.38 &    0.23 &      0.01 --  0.36 \\
ESO 286-IG 019            &    0.0430 &   43.38 &          $<$ 43.40 &   43.83 &          $<$ 43.84 &   -1.80 &     -1.97 -- -1.73 &   -3.99 &     -4.17 -- -3.93 \\
ESO 297- G 018            &    0.0252 &   42.51 &     42.37 -- 42.62 &   43.01 &     42.95 -- 43.06 &   -2.20 &     -2.61 -- -1.91 &   -0.17 &     -0.41 -- -0.03 \\
\enddata
\tablecomments{Table 2 is published in its entirety in the electronic edition of the Astrophysical Journal. A portion is shown here for guidance regarding its form and content.}
\end{deluxetable*}

\end{document}